\documentclass[aps,showpacs,amsfonts,nofootinbib,preprintnumbers,nobalancelastpage]{revtex4}

\usepackage{graphicx}
\usepackage{epsfig}
\usepackage{xcolor}
\usepackage{rotating}
\usepackage{amsmath}
\usepackage{braket}

\newcommand{\cw}{{\rm c_W} }
\newcommand{\sw}{{\rm s_W} }

\newcommand{\cwsq}{{\rm c_W^2} }
\newcommand{\swsq}{{\rm s_W^2} }

\newcommand{\st}{\rm s_W}
\newcommand{\ct}{\rm c_W}
\newcommand{\cdt}{\rm c_{2\theta_W}}
\newcommand{\sdt}{\rm s_{2\theta_W}}

\newcommand{\na}{{\rm \--}}
\newcommand{\nc}{{{\rm N}_{\rm c}}}
\newcommand{\Ng}{{{\rm N}_{\rm g}}}

\newcommand{\Dfb}{\mbox{$\raisebox{2mm}{\boldmath ${}^\leftrightarrow$}
\hspace{-4mm} D$}}
\newcommand{\Dfba}{\mbox{$\raisebox{2mm}{\boldmath ${}^\leftrightarrow$}\hspace{-4mm} D^a$}}

\newcommand {\fbw}     {f_{BW}}
\newcommand {\fpone}   {f_{\Phi,1}}

\newcommand {\fw}      {f_{W}}
\newcommand {\fb}      {f_{B}}

\newcommand {\opone}   {{\cal O}_{\Phi,1}}

\preprint{\bf YITP-SB-17-19}

\begin{document}
\title{Unitarity Constraints on Dimension-six Operators II:
  Including Fermionic Operators}

\author{Tyler Corbett}
\email{corbett.t.s@gmail.com}
\affiliation{ARC Centre of Excellence for Particle Physics at the Terascale,
School of Physics, The University of Melbourne, Victoria, Australia}

\author{O.\ J.\ P.\ \'Eboli}
\email{eboli@fma.if.usp.br}
\affiliation{Instituto de F\'{\i}sica,
             Universidade de S\~ao Paulo, S\~ao Paulo -- SP, Brazil}

\author{M.\ C.\ Gonzalez--Garcia} \email{concha@insti.physics.sunysb.edu}
\affiliation{%
  Instituci\'o Catalana de Recerca i Estudis Avan\c{c}ats (ICREA),}
\affiliation {Departament d'Estructura i Constituents de la Mat\`eria, 
Universitat
  de Barcelona, 647 Diagonal, E-08028 Barcelona, Spain}
\affiliation{%
  C.N.~Yang Institute for Theoretical Physics, SUNY at Stony Brook,
  Stony Brook, NY 11794-3840, USA}

\begin{abstract}

  We analyze the scattering of fermions, Higgs and electroweak gauge
  bosons in order to obtain the partial--wave unitarity bounds on
  dimension--six effective operators, including those involving
  fermions.  We also quantify whether, at the LHC energies, the
  dimension--six operators lead to unitarity violation after taking
  into account the presently available constraints on their Wilson
  coefficients.  Our results show that for most dimension--six
  operators relevant for the LHC physics there is no unitarity
  violation at the LHC energies, and consequently, there is no need
  for the introduction of form factors in the experimental and
  phenomenological analyses, making them model independent. We also
  identify two operators for which unitarity violation is still
  an issue at the LHC Run--II.

\end{abstract}


\maketitle
\renewcommand{\baselinestretch}{1.15}
%

\section{Introduction}

The discovery of a scalar state with properties in agreement with
those of the Standard Model (SM) Higgs boson at the Large Hadron
Collider (LHC) set the final stone in establishing the validity of the
model.  Presently there are no high energy data that are in
significant conflict with the SM predictions.  In this framework, with
no other new state yet observed, one can parametrize generic
departures from the SM by an effective Lagrangian constructed with the
SM states and respecting the SM symmetries, abandoning only the
renormalizabilty condition which constrains the dimension of the
operators to be of dimension four or less. In particular the
established existence of a particle resembling the SM Higgs boson
implies that the $SU(3)_C \otimes SU(2)_L \otimes U(1)_Y$ symmetry can
be realized linearly in the effective theory, an assumption under
which we will work in this paper.  In this framework the first
departures from the SM at the LHC which also respect its global
symmetries appear at dimension--six. \smallskip

When using such an effective field theory to quantify possible
deviations from the SM predictions, one must be sure of its validity
in the energy range probed by the experiments.  As is well known,
the higher--dimensional operators included in the effective Lagrangian
can lead to perturbative partial--wave unitarity violation at high
energies, signaling a maximum value of the center--of--mass energy for
its applicability.  If there is unitarity violation we must either
modify the effective field theory, {\em e.g.}  by adding form factors
that take into account higher order terms~\cite{Baur:1987mt}, or it
should be replaced by an ultraviolet (UV) complete model. In
Ref.~\cite{Corbett:2014ora} we presented a general study of unitarity
violation in electroweak and/or Higgs boson pair production in boson
and/or fermion collisions associated with the presence of
dimension--six operators involving bosons, concentrating on those which
are blind to low energy bounds.  The rationale behind this choice was
that not--blind operators were expected to be too constrained by
electroweak precision data to be relevant.  In this work we revisit
this assumption and extend the study to introduce the effects of these
operators sensitive to low energy observables, and in particular those
involving the coupling of fermions to electroweak bosons.  This is
timely since the LHC has already started to be able to probe triple
electroweak gauge boson couplings with a precision comparable to, or
even better than, LEP~\cite{Butter:2016cvz}. With such a precision the
LHC experiments are already sensitive to deviations of the couplings
of electroweak gauge bosons to fermions that are of the order of the
limits obtained using the electroweak precision data (EWPD)
~\cite{Farina:2016rws, Zhang:2016zsp}. \smallskip

In this paper we evaluate the unitarity bounds on bosonic and
fermionic dimension--six operators from boson pair production
amplitudes.  As in Ref.~\cite{Corbett:2014ora} we take into account
all coupled channels in both elastic and inelastic scattering and all
possible helicity amplitudes. Moreover, we consistently work at fixed
order in the effective Lagrangian expansion\footnote{Other studies in the literature have been performed either
considering only one
non--vanishing coupling at a time, and/or they did not take into
account coupled channels, or they worked in the framework of effective
vertices~\cite{Bilchak:1987cp, Gounaris:1993fh, Gounaris:1994cm,
Gounaris:1995ed, Degrande:2013mh,Baur:1987mt,Dahiya:2013uba,Ghosh:2017coz}
}.

We also study the variation of the constraints
under the assumption of the flavour dependence of the fermionic
operators.  \smallskip

With these results we can address whether, at the LHC energies, the
dimension--six operators can indeed lead to unitarity violation after
taking into account the presently available constraints on the
anomalous couplings. This is accomplished by substituting the
present limits of the Wilson coefficients in our partial--wave
unitarity bounds to extract the center--of--mass energy at which
perturbative unitarity is violated.  In order to do so we consistently
derive the EWPD constraints on the coefficients of the non-blind
operators.
Our results show that for all dimension--six operators relevant
  for the LHC physics, except for just two (${\cal O}_{\Phi,2}$ and
  ${\cal O}^{(1)}_{\Phi d}$), there is no unitarity violation at the
  LHC energies, and consequently, we can safely neglect the
  introduction of form factors in the experimental and
  phenomenological analyses, making them cleaner and free of ad-hoc
  parameters. In the case of the operator ${\cal O}_{\Phi,2}$ there is
  no unitarity violation up to subprocess center--of--mass energies of
  the order of 2.1 TeV, meaning that we have to be more careful in
  analyzing the high energy tail of processes where the Higgs boson
  can participate.  On the other hand for ${\cal O}^{(1)}_{\Phi d}$
  perturbative unitarity holds for diboson (VV) subprocess
  center--of--mass energy less than 3.5 TeV. \smallskip

This paper is organized as follows. Section~\ref{sec:lag} contains the
dimension--six operators relevant for our analyses, while we present
in Section~\ref{sec:unitbounds} the unitarity bounds for bosonic and
fermionic operators (listing the unitarity violating amplitudes in
Appendix~\ref{app:helamp}).  We discuss the consequences of these
results in Section~\ref{sec:disc} taking into account the existing
constraints on the Wilson coefficients of the dimension--six
operators.  We present the details of our fit to the EWPD in
Appendix~\ref{app:ewpd}, while for completeness we summarize in
Appendix~\ref{app:dipole} the unitarity constraints on fermion dipole
operators.

\section{Effective Lagrangian}
\label{sec:lag}

We parametrize deviations from the Standard Model (SM) in terms of
higher dimension operators as
\begin{equation}
   {\cal L}_{\rm eff} = {\cal L}_{\rm SM} + \sum_{n>4,j}
   \frac{f_{n,j}}{\Lambda^{n-4}} {\cal O}_{n,j} \;.
\end{equation}
The first operators that impact the LHC physics are of $n=6$, or
dimension--six.  Their basis contains 59 independent operators, up to
flavor and hermitian conjugation, where we impose the SM gauge
symmetry, as well as baryon and lepton number
conservation~\cite{Buchmuller:1985jz,Grzadkowski:2010es}. Of those, 49
can be chosen to be C and P conserving and do not involve gluons.
Since the S-matrix elements are unchanged by the use of the equations
of motion (EOM), there is a freedom in the choice of basis~\cite{
  Politzer:1980me, Georgi:1991ch,Arzt:1993gz,Simma:1993ky}. Here we
work in that of Hagiwara, Ishihara, Szalapski, and
Zeppenfeld~\cite{Hagiwara:1993ck, Hagiwara:1996kf}.

\subsection{Bosonic Operators} 

Assuming $C$ and $P$ conservation there are nine 
dimension--six operators in our basis involving only bosons
that take part at tree level in
two--to--two scattering of gauge and Higgs bosons after we employ the
EOM to eliminate redundant operators~\cite{Corbett:2012ja}.  We group
these operators according to their field content. In the first class
there is just one operator that contains exclusively gauge bosons.
\begin{equation}
  \mathcal{O}_{WWW} = {\rm Tr}[\widehat{W}_{\mu}^{\nu}
    \widehat{W}_{\nu}^{\rho}\widehat{W}_{\rho}^{\mu}] \;\;.
\label{eq:www}
\end{equation}
In the next group there are six operators that include Higgs
and electroweak gauge fields.
\begin{equation}
\begin{array}{l@{\hspace{1cm}}l}
  \mathcal{O}_{WW}		
= \Phi^\dagger\widehat{W}_{\mu\nu}\widehat{W}^{\mu\nu}\Phi \;\;, 
  &\mathcal{O}_{BB}		
= \Phi^\dagger\widehat{B}_{\mu\nu}\widehat{B}^{\mu\nu}\Phi  \;\;,	\\
  \mathcal{O}_{BW}		
= \Phi^\dagger\widehat{B}_{\mu\nu}\widehat{W}^{\mu\nu}\Phi  \;\;,	
  &\mathcal{O}_{\Phi,1}		
=	(D_\mu\Phi)^\dagger\Phi\Phi^\dagger(D^\mu\Phi) \;\;,	
\\
 \mathcal{O}_{W}		
=	(D_\mu\Phi)^\dagger\widehat{W}^{\mu\nu}(D_\nu\Phi)  \;\;,	
  &\mathcal{O}_{B}		
=	(D_\mu\Phi)^\dagger\widehat{B}^{\mu\nu}(D_\nu\Phi)  \;\;.	
\end{array}
\end{equation}
The final class contains two operators expressed solely in terms
of Higgs fields
\begin{equation}
  \mathcal{O}_{\Phi,2} =
  \frac{1}{2}\partial^\mu(\Phi^\dagger\Phi)\partial_\mu(\Phi^\dagger\Phi)
                             \;\; \hbox{ and } \;\;
  \mathcal{O}_{\Phi,3} =\frac{1}{3}(\Phi^\dagger\Phi)^3  \;\;.
\label{eq:phi3}
\end{equation}
Here $\Phi$ stands for the Higgs doublet and we have adopted the notation
$\widehat{B}_{\mu\nu} \equiv i(g^\prime/2)B_{\mu\nu}$,
$\widehat{W}_{\mu\nu} \equiv i(g/2)\sigma^aW^a_{\mu\nu}$, with $g$ and
$g^\prime$ being the $SU(2)_L$ and $U(1)_Y$ gauge couplings
respectively, and $\sigma^a$ the Pauli matrices. \smallskip

The dimension--six operators given in
Eqs.~(\ref{eq:www})--(\ref{eq:phi3}) affect the scatterings
$VV \to VV$ and $f \bar{f} \to VV$, where $V$ stands for the
electroweak gauge bosons or the Higgs, through modifications of triple
and quartic gauge boson couplings, Higgs couplings to fermions and
gauge bosons, interactions of gauge bosons with fermion pairs, and the
Higgs self-couplings; see Table~\ref{tab:coupl}.  Moreover, these
anomalous couplings enter in the analyses of Higgs physics, as well
as, triple gauge couplings of electroweak gauge bosons and were
analyzed in~\cite{Corbett:2012dm, Corbett:2012ja, Corbett:2013pja,
  Corbett:2015ksa, Corbett:2014ora}. \smallskip

\subsection{Operators with fermions}

After requiring that the dimension--six operators containing
fermions conserve $C$, $P$ and baryon number, we are left with 40
independent operators (barring flavour indexes) in our basis which
do not involve gluon fields. We classify them in
four groups. In the first group there are three dimension--six
operators that modify the Yukawa couplings of the Higgs boson, and
therefore do not contribute to the processes that we study at high
energies. The second class possesses 25 four--fermion contact
interactions that again do not take part in our analyses. \smallskip

The third group includes the operators that lead to anomalous
couplings of the gauge bosons with the fermions
that exhibit the same Lorentz structures as the SM vertices
and are relevant for our analyses. This class contains eight
dimension--six operators
\begin{equation}
  \begin{array}{l@{\hspace{1cm}}l@{\hspace{1cm}}l}
& 
{\cal O}^{(1)}_{\Phi L,ij}=\Phi^\dagger (i\, \Dfb_\mu \Phi) 
(\bar L_{i}\gamma^\mu L_{j}) ,
& 
{\cal O}^{(3)}_{\Phi L,ij}
=\Phi^\dagger (i\,{\Dfba}_{\!\!\mu} \Phi) 
(\bar L_{i}\gamma^\mu T_a L_{j}) , \\
& 
{\cal O}^{(1)}_{\Phi Q,ij}=\Phi^\dagger (i\,\Dfb_\mu \Phi)  
(\bar Q_i\gamma^\mu Q_{j}) ,
& 
{\cal O}^{(3)}_{\Phi Q,ij}=\Phi^\dagger (i\,{\Dfba}_{\!\!\mu} \Phi) 
(\bar Q_i\gamma^\mu T_a Q_j) ,\\
& 
{\cal O}^{(1)}_{\Phi e,ij}=\Phi^\dagger (i\Dfb_\mu \Phi) 
(\bar e_{R_i}\gamma^\mu e_{R_j})  ,
& 
\\
& {\cal O}^{(1)}_{\Phi u,ij}=\Phi^\dagger (i\,\Dfb_\mu \Phi) 
(\bar u_{R_i}\gamma^\mu u_{R_j}) ,
& \\
& {\cal O}^{(1)}_{\Phi d,ij}=\Phi^\dagger (i\,\Dfb_\mu \Phi) 
(\bar d_{R_i}\gamma^\mu d_{R_j}) ,
& \\

& {\cal O}^{(1)}_{\Phi ud,ij}=\tilde\Phi^\dagger (i\,\Dfb_\mu \Phi) 
(\bar u_{R_i}\gamma^\mu d_{R_j} +{\rm h.c.}) ,
& 
\end{array}
\label{eq:hffop}
\end{equation}
where we defined $\tilde \Phi=i \sigma_2\Phi^*$,
$\Phi^\dagger\Dfb_\mu\Phi= \Phi^\dagger D_\mu\Phi-(D_\mu\Phi)^\dagger
\Phi$ and $\Phi^\dagger \Dfba_{\!\!\mu} \Phi= \Phi^\dagger T^a D_\mu
\Phi-(D_\mu\Phi)^\dagger T^a \Phi$
where $T^a=\sigma^a/2$.  We have also used the notation of $L$ for the
lepton doublet, $Q$ for the quark doublet and $f_R$ for the $SU(2)_L$
singlet fermions, where $i, j$ are family indices. \smallskip

The set of operators in Eq.~(\ref{eq:hffop}) is redundant as two can
be removed by the use of the EOM of the electroweak gauge bosons.
We chose to remove from the
basis  the following combinations of fermionic
operators~\cite{Corbett:2012ja}\footnote{This is a different choice with
respect to the basis in Ref.~\cite{Grzadkowski:2010es},
there these two fermionic
operators are kept in exchange of the bosonic operators ${\cal O}_W$
and ${\cal O}_B$.}
\begin{equation}
  \sum_i {\cal O}^{(1)}_{\Phi L,ii} , \;\;\;{\rm and}\;\;\;
  \sum_i {\cal O}^{(3)}_{\Phi L,ii} \;.
\label{eq:EOMred}  
\end{equation}

We notice that the operators in the third group not only contribute to
$VV \to VV$ and $ f \bar{f} \to VV$ processes, but they can also be
bounded by the EWPD, in particular from $Z$--pole and $W$--pole
observables; see Section~\ref{sec:disc} and
Appendix~\ref{app:ewpd}. By using the EOM to remove the combinations
in Eq.~(\ref{eq:EOMred}) we have selected the operator basis in such a
way that there are no blind directions in the analysis of the EWPD
data~\cite{Corbett:2012ja}.  \smallskip

To avoid the generation of too large flavor violation, in what follows we
assume no generation mixing in these operators, that is, for any
operator ${\cal O}_{ij}={\cal O}_{ii}\delta_{ij}$. \smallskip

Finally we notice that the complete basis of dimension--six operators
also contains a fourth group of dipole fermionic operators ({\em i.e.}
with tensor Lorentz structure) and that can participate in two--to--two
scatterings of fermions into gauge and Higgs bosons but that do not
modify the $Z$--pole and $W$--pole physics at tree level,
since their contributions do not interfere with the SM ones. They are
\begin{equation}
\begin{array}{l@{\hspace{1cm}}l@{\hspace{1cm}}l}
{\cal O}_{eW,ij} = i \bar{L_i} \sigma^{\mu\nu} \ell_{R,j} \widehat{W}_{\mu\nu}
  \Phi \;\;\;, 
& {\cal O}_{eB,ij} = i \bar{L_i} \sigma^{\mu\nu} \ell_{R,j} \widehat{B}_{\mu\nu}
  \Phi \;\;\;, 
\\
{\cal O}_{uW,ij} = i \bar{Q_i} \sigma^{\mu\nu} u_{R,j} \widehat{W}_{\mu\nu}
  \tilde\Phi \;\;\;, 
&{\cal O}_{uB,ij} = i \bar{Q_i} \sigma^{\mu\nu} u_{R,j} \widehat{B}_{\mu\nu}
  \tilde\Phi \;\;\;, 
\\
{\cal O}_{dW,ij} =  i \bar{Q_i} \sigma^{\mu\nu} d_{R,j} \widehat{W}_{\mu\nu}
  \Phi \;\;\;, 
& {\cal O}_{dB,ij} =  i \bar{Q_i} \sigma^{\mu\nu} u_{R,j} \widehat{B}_{\mu\nu}
  \Phi \;\;\;,
\end{array}
\label{eq:dipole}
\end{equation}
where $i$,$j$ are family indices. These operators lead to partial--wave unitarity
violation in different channels from the operators in Eq.~(\ref{eq:hffop}), and
therefore can be bounded independently. For completeness
we present the corresponding unitarity violating amplitudes and bounds
in Appendix \ref{app:dipole}. \smallskip

\begin{table}
{\footnotesize  \begin{tabular}{|c|c|c|c|c|c|c|c|c|c|c|c|c|}
\hline 
& $VVV$ & $VVVV$ & $HVV$ & $HVVV$ & $HHVV$ &
$HHH$ & $HHHH$ & $H\bar{f}f$  & $Z\bar{q}{q}$
& $Z\bar{l}{l}$   & $W\bar{u}{d}$   & $W\bar{l}{\nu}$   
\\ \hline 
$\mathcal{O}_{WWW}$ & X & X & & & & & &  & &  & &
\\ \hline 
$\mathcal{O}_{WW}$ & & & X & X & X & & &  & &  & &
\\ \hline
$\mathcal{O}_{BB}$ & & & X &   & X & & &  & &  & &
\\ \hline 
$\mathcal{O}_{BW}$ & X & X & X & X & X & & &  & X & X & X & {X}
\\ \hline 
$\mathcal{O}_{W}$ & X & X & X & X & X & & &  & &  & &
\\ \hline 
$\mathcal{O}_{B}$ & X & & X & X & X & & &  & &  & &
\\ \hline
${\mathcal{O}_{\Phi,1}}$ & X  &X & X & & X & X & X & X & {X} & {X} & {X} &{X} 
\\ \hline
$\mathcal{O}_{\Phi,2}$ & & & X & & X & X & X & X  & &  & &
\\ \hline
$\mathcal{O}_{\Phi,3}$ & & & & & & X & X &  & &  & &
\\ \hline
${\cal O}^{(1)}_{\Phi Q}$,
${\cal O}^{(1)}_{\Phi u}$,
${\cal O}^{(1)}_{\Phi d}$ 
& & &   & &   &   &   &   & X  &  & &
 \\ \hline 
${\cal O}^{(3)}_{\Phi Q}$,
& & &   & &   &   &   &   & X  &  & X &
\\ \hline 
${\cal O}^{(1)}_{\Phi ud}$
& & &   & &   &   &   &   &   &  & X&
\\\hline
${\cal O}^{(1)}_{\Phi L}$,${\cal O}^{(1)}_{\Phi e}$,
& & &   & &   &   &   &   &   & X & &
\\ \hline 
${\cal O}^{(3)}_{\Phi L}$,
& & &   & &   &   &   &   & X  &  & X & X
\\ \hline 
\end{tabular}
\caption{Couplings relevant for our analysis that are modified by the
dimension--six operators in Eqs.~(\ref{eq:www})--(\ref{eq:hffop}). 
Here, $V$ stands for any   electroweak gauge boson, $H$ for the Higgs 
and $f$ for SM fermions.}
\label{tab:coupl}
}
\end{table}

\section{Constraints from Unitarity Violation in two--to--two Processes }
\label{sec:unitbounds}

Let us start by studying the unitarity violating amplitudes associated
with the bosonic operators listed in
Eqs.~(\ref{eq:www})--(\ref{eq:phi3}) of
which all but ${\cal O}_{\Phi,3}$ lead to amplitudes which grow with
$s$ in the two--to--two scattering of electroweak gauge bosons and Higgs
\begin{equation}
{V_1}_{\lambda_1}{V_2 }_{\lambda_2} \to {V_3}_{\lambda_3}{V_4}_{\lambda_4} \;\;.
\end{equation}
The  helicity amplitude of these processes is then expanded in partial waves
in the center--of--mass system, following the conventions of~\cite{Jacob:1959at}
\begin{equation}
\mathcal{M}
({V_1}_{\lambda_1}{V_2 }_{\lambda_2} \to {V_3}_{\lambda_3}{V_4}_{\lambda_4}) 
=16 \pi \sum_J
\left ( J+\frac{1}{2} \right)~ 
\sqrt{1+\delta_{{V_1}_{\lambda_1}}^{{V_2}_{\lambda_2}}}
\sqrt{1+\delta_{{V_3}_{\lambda_3}}^{{V_4}_{\lambda_4}}}
d_{\lambda\mu}^{J}(\theta) ~e^{i M \varphi}
~ T^J({V_1}_{\lambda_1}{V_2 }_{\lambda_2} \to {V_3}_{\lambda_3}{V_4}_{\lambda_4}) 
\;\;,
\label{eq:helamp}
\end{equation}
where $d$ is the usual Wigner rotation matrix and
$\lambda=\lambda_1-\lambda_2$, $\mu=\lambda_3-\lambda_4$,
$M = \lambda_1 - \lambda_2 - \lambda_3 + \lambda_4$, and $\theta$
($\varphi$) is the polar (azimuth) scattering angle. In the case one
of the vector bosons is replaced by the Higgs we use this expression
by setting the correspondent $\lambda$ to zero.
\smallskip

The helicity scattering amplitudes for the operators $\opone$ and
${\cal O}_{BW}$ are presented in the Appendix~\ref{app:helamp}, while
the corresponding amplitudes for the other bosonic operators can be
found in Ref.~\cite{Corbett:2014ora}.  Notice that the contributions
of the bosonic operators to $VV \to VV$ scattering amplitudes grow
with $s$ since the gauge invariance leads to the cancellation of
potential terms growing as $s^2$~\cite{Csaki:2003dt}.  Moreover, all
bosonic operators contribute to the $J=0$ and $J=1$ partial waves,
however, ${\cal O}_{WWW}$ also leads to the growth of $J \ge 2$
amplitudes. Nevertheless, the most stringent bounds come from the
$J=0$ and $1$ partial waves, therefore, we restrict our attention to
these channels. Furthermore unitarity violating amplitudes arise for
the three possible charge channels $Q=0,1,2$; see
Ref.~\cite{Corbett:2014ora} for notation and a list of all the states
contribution to each $(Q,J)$ channel. \smallskip

In order to obtain the strongest bounds on the coefficients of the
eight operators, we diagonalize the six matrices containing the
$T^J_Q$ amplitudes for each of the $(Q,J)$ channels and impose that
all their eigenvalues (a total of 59) satisfy the constraint
\begin{equation}
|T^J({V_1}_{\lambda_1}{V_2 }_{\lambda_2} \to {V_1}_{\lambda_1}{V_2}_{\lambda_2}) 
| \le 2 \;\;.
\label{eq:unitcond}
\end{equation}

Initially we obtain the unitarity bounds on the eight bosonic
operators assuming that only one Wilson coefficient differs from zero
at a time.  This is a conservative scenario, {\em i.e.} leads to
stringent bounds, since we do not take into account that more than one
operator contributing to a given channel could lead to cancellations
and therefore looser limits. For this case we obtain:
\begin{equation}
\begin{array}{l@{\hspace{2cm}}l@{\hspace{2cm}}l}

\left|\frac{f_{\phi,2}}{\Lambda^2}s\right|\le33 \;\;\;,\;\;\;
&\left|\frac{f_{\phi,1}}{\Lambda^2}s\right|\le50 \;\;\;,\;\;\;
&\left|\frac{f_W}{\Lambda^2}s\right| \le 87 \;\;\;,\;\;\;
\\
&&
\\
\left|\frac{f_B}{\Lambda^2}s\right| \le 617 \;\;\;,\;\;\;
& \left|\frac{f_{WW}}{\Lambda^2}s\right| \le 99 \;\;\;,\;\;\;
& \left|\frac{f_{BB}}{\Lambda^2}s\right| \le 603 \;\;\;,\;\;\;
\\
&&
\\
  \left|\frac{f_{BW}}{\Lambda^2}s\right| \le 456 \;\;\;,\;\;\;
&\left|\frac{f_{WWW}}{\Lambda^2}s\right| \le 85 \;\;\;.
\end{array}
\label{eq:bound1vvvv}
\end{equation}

Next we study the constraints on the full set of eight bosonic
operators when they are all allowed to vary. In order to find closed
ranges in the eight--dimensional parameter space we need to consider
also the constraints from fermion annihilation into electroweak gauge
bosons~\cite{Baur:1987mt}.  To do so we obtain the helicity
amplitudes of all processes
\begin{equation}
f_{1\sigma_1} \bar{f}_{2\sigma_2} \to V_{3\lambda_3} V_{4\lambda_4}\;\;\; ,
\label{eq:fscat}
\end{equation}
and then perform the expansion in partial waves of the
center--of--mass system; for further details and conventions see
Ref.~\cite{Corbett:2014ora}. \smallskip

We present in the Appendix~\ref{app:helamp} the leading order terms of
the scattering amplitudes that give rise to unitarity violation at
high energies taking into account the dimension--six operators in
Eqs.~(\ref{eq:www})--(\ref{eq:hffop}).  These amplitudes are
proportional to $\delta_{\sigma_1,-\sigma_2}$ since we neglect the
fermion masses in the high energy limit.  It is interesting to notice
that the dimension--six operators ${\cal O}_{WWW}$, ${\cal O}_W$ and
${\cal O}_B$ modify the triple electroweak gauge boson couplings
(TGC), therefore, as expected, their presence would affect the SM
cancellations that cut off the growth of the $f \bar{f} \to VV$
amplitudes. On the other hand, the operator ${\cal O}_{BW}$ also
modifies the TGC, however, its effects on the $Z/\gamma$ wave function
renormalizations cancel the growth with the center--of--mass energy
due to the anomalous TGC. Similar cancellations occur for
  ${\cal O}_{\Phi,1}$ which contributes to triple gauge vertices, as
  well as the coupling of gauge bosons to fermions. \smallskip

In order to obtain more stringent bounds and to separate the
contributions of the different operators, we consider
the processes~\cite{Baur:1987mt}
\begin{equation}
 X \to V_{3\lambda_3} V_{4\lambda_4} \;\;\; ,
\end{equation}
where $X$ is a linear combination of fermionic initial states:
\begin{equation}
   | X \rangle = \sum_{f_i \sigma_i}
x_{f_1 \sigma_1, f_2\sigma_2} | f_{1\sigma_1} \bar{f}_{2\sigma_2} \rangle \;,
\end{equation}
with the normalization $  \sum_{f_i \sigma_i}
| x_{f_1 \sigma_1, f_2\sigma_2} |^2 =1$.
The corresponding bounds read~\cite{Baur:1987mt}
\begin{equation}
\sum_{{V_3}_{\lambda_3},{V_4}_{\lambda_4}}
\left|T^J(X\to 
{V_3}_{\lambda_3}{V_4}_{\lambda_4})\right|^2 \leq 1 \; . 
\label{eq:unitcond2}
\end{equation}

In particular using the linear combinations as displayed in the first
three lines of Table~\ref{tab:fernogen} we are able to impose
independent bounds in each of the Wilson coefficients of the three
bosonic operators participating in the $f\bar{f} \rightarrow VV$
scattering amplitudes. \smallskip

\begin{table}
\scalebox{0.7}{
\begin{tabular}{l|l|l}
  Fermion State &   $V_3 V_4$ &$T^{J=1}$  \\\hline
$\frac{1}{\sqrt{2\Ng(1+\nc)}}|{\displaystyle \sum_{i=1}^{\Ng}}
\left(\,-\,e^-_{-,i}\,e^+_{+,i}\,+\,\,\nu_{-,i}\bar\nu_{+,i}\, +
{\displaystyle\sum_{a=1}^{\nc}} \left(\,-\,d^a_{-,i}\,\bar d^a_{+,i}\,
+\, u^a_{-,i}\,\bar u^a_{+,i}\right) \right)\rangle$
& $W^+_- W^-_-$, $W^+_+ W^-_+$ &
${+}i \frac{1}{12\sqrt{2}\pi} \sqrt{2\Ng(\nc+1)}
\frac{3g^4}{4} \frac{f_{WWW}\, s }{\Lambda^2}$
\\ \hline
$\frac{1}{\sqrt{2 \Ng}}
  |{\displaystyle \sum_{i=1}^{\Ng}}
\left(e^-_{-,i}\,e^+_{+,i}\,+\,\nu_{-,i}\bar\nu_{+,i}\right)\rangle$
& $W^+_0 W^-_0$ &
${+}i \frac{1}{12\sqrt{2}\pi} \sqrt{\frac{\Ng}{2}}
\frac{g^2 \sw^2}{4\,\cw^2}  \frac{f_B \, s}{\Lambda^2}$
\\ \hline
$\frac{1}{\sqrt{2 \Ng}}|\displaystyle{\sum_{i=1}^{\Ng}}
  (\left(e^-_{-,i}\,e^+_{+,i}\,-\,\nu_{-,i}\bar\nu_{+,i}\right)\rangle$
  &
$W^+_0 W^-_0$ &
${+}i \frac{1}{12\sqrt{2}\pi} \sqrt{\frac{\Ng}{2}}
\frac{g^2}{4}  \frac{f_W \, s}{\Lambda^2}$
\\\hline
$\frac{1}{\sqrt{3 \Ng}}|{\displaystyle \sum_{i=1}^{\Ng}}
\left(e^-_{-,i}\,e^+_{+,i}\,+\,\nu_{-,i}\bar\nu_{+,i}\, {+}\,e^-_{+,i}\,e^+_{-,i}\right)\rangle$
&
$W^+_0 W^-_0$ &
${-} i \frac{1}{12\sqrt{2}\pi} \sqrt{\frac{\Ng}{3}}
\frac{f^{(1)}_{\Phi e} \, s}{\Lambda^2}$
\\\hline
$\frac{1}{\sqrt{\Ng(\nc+8)}}|{\displaystyle \sum_{i=1}^{\Ng}}
\left(2\,e^-_{-,i}\,e^+_{+,i}\,+2\,\,\nu_{-,i}\bar\nu_{+,i}\, {-}
     {\displaystyle\sum_{a=1}^{\nc}} u^a_{+,i}\,\bar u^a_{-,i}\right)\rangle$
&
$W^+_0 W^-_0$ &
$ {+}i \frac{1}{12\sqrt{2}\pi} \frac{1}{\sqrt{\Ng(\nc+8)}} \Ng \nc
\frac{f^{(1)}_{\Phi u} \, s}{\Lambda^2}$
\\\hline
$\frac{1}{\sqrt{\Ng(\nc+2)}}
|{\displaystyle \sum_{i=1}^{\Ng}}
\left(\,e^-_{-,i}\,e^+_{+,i}\,+\,\,\nu_{-,i}\bar\nu_{+,i}\, {+}
     {\displaystyle\sum_{a=1}^{\nc}} d^a_{+,i}\,\bar d^a_{-,i}\right)\rangle$
& $W^+_0 W^-_0$ &
$ {-}i \frac{1}{12\sqrt{2}\pi} \frac{\Ng \nc}{\sqrt{\Ng(\nc+2)}} 
\frac{f^{(1)}_{\Phi d} \, s}{\Lambda^2}$
\\\hline
$\frac{1}{\sqrt{\Ng(2\nc+2)}}|{\displaystyle \sum_{i=1}^{\Ng}}
\left(\,e^-_{-,i}\,e^+_{+,i}\,+\,\,\nu_{-,i}\bar\nu_{+,i}\, +
{\displaystyle\sum_{a=1}^{\nc}} \left(d^a_{-,i}\,\bar d^a_{+,i}\,
+\, u^a_{-,i}\,\bar u^a_{+,i}\right) \right)\rangle$
& $W^+_0 W^-_0$ &
$ {+}i \frac{1}{12\sqrt{2}\pi} \frac{2 \Ng \nc}{\sqrt{\Ng(2\nc+2)}} 
\frac{f^{(1)}_{\Phi Q} \, s}{\Lambda^2}$
\\\hline
$\frac{1}{\sqrt{4\Ng }}
 |{\displaystyle \sum_{i=1}^{\Ng}}
\left(\,e^-_{-,i}\,e^+_{+,i}\,-\,\,\nu_{-,i}\bar\nu_{+,i}\, +
u^a_{-,i}\,\bar u^a_{+,i} -d^a_{-,i}\,\bar d^a_{+,i}
\right)\rangle$
& $W^+_0 W^-_0$ &
${+} i \frac{1}{12\sqrt{2}\pi} \frac{\Ng}{\sqrt{4}}  \frac{1}{2}
\frac{f^{(3)}_{\Phi Q} \, s}{\Lambda^2}$
\\\hline
$\frac{1}{\sqrt{\Ng\nc}}
  |{\displaystyle \sum_{i=1}^{\Ng}}{\displaystyle \sum_{a=1}^{\nc}}
  d^a_{+,i}\,\bar u^a_{-,i} \rangle$
&  $W^+_0 W^-_0$ &
$ -i \frac{1}{12\sqrt{2}\pi} \sqrt{\Ng\nc} \sqrt{2}
\frac{f^{(1)}_{\Phi ud} \, s}{\Lambda^2}$
\\\hline
\end{tabular}  }
\caption{Initial fermionic states used to obtain bounds on the
  generation independent fermionic operators.  We also present the
  high--energy dominant terms of the corresponding amplitudes.
  We denote by $\Ng=3$
  the number of generations while $\nc=3$ is the number of
  colors. $\sqrt{s}$ stands for the center--of--mass energy of the processes.}
\label{tab:fernogen}
\end{table}

Combining those with the conditions from partial--wave unitarity of the
59 eigenvalues of the elastic boson scattering amplitudes discussed
above we find the most general constraints in the eight--dimensional
parameter space. In summary, allowing all coefficients to be nonzero,
and searching for the largest allowed values for each operator
coefficient while varying over the other coefficients, yields:
\begin{equation}
\begin{array}{l@{\hspace{2cm}}l@{\hspace{2cm}}l}
\left|\frac{f_{\phi,2}}{\Lambda^2}s\right| \le 209 \;\;\;,\;\;\;
&\left|\frac{f_{\phi,1}}{\Lambda^2}s\right| \le 151 \;\;\;,\;\;\;
&\left|\frac{f_W}{\Lambda^2}s\right| \le 436 \;\;\;,\;\;\;
\\
\left|\frac{f_B}{\Lambda^2}s\right| \le 1460 \;\;\;,\;\;\;
&\left|\frac{f_{WW} }{\Lambda^2}s\right| \le 319 \;\;\;,\;\;\;
& \left|\frac{f_{BB} }{\Lambda^2}s\right| \le 1340 \;\;\;,\;\;\;
\\
\left|\frac{f_{BW}}{\Lambda^2}s\right| \le 1386 \;\;\;,\;\;\;
&\left|\frac{f_{WWW}}{\Lambda^2}s\right| \le {85} \;\;\;.\;\;\;
\end{array}
\label{eq:boundMvvvv}
\end{equation}
As expected these bounds extend the region of validity of the
effective theory with respect to the case where just one Wilson
coefficient is allowed to be non--vanishing. This is the most
optimistic scenario because it implicitly assumes that the values of
the Wilson coefficients are tuned to have the largest energy region
where the effective theory is valid. \smallskip

It is important to stress that both the one--dimensional bounds in
Eq.~\eqref{eq:bound1vvvv} and the general bounds in
Eq.~\eqref{eq:boundMvvvv} hold independently of the values of the
coefficients of the fermionic operators due to the choice of initial
states in Table~\ref{tab:fernogen}.

\subsection{Bounds on Generation Independent Operators}

Now we focus our attention on the operators involving fermions, what
require assumptions concerning their flavour structure as we discuss
next.  Initially we assume that the new physics giving rise to the
dimension--six operators is generation blind.  In this case we can
drop the generation index in all coefficients.  Therefore, the
constraint on the operators in Eq.~(\ref{eq:EOMred}) implies that the
operators ${\cal O}_{\Phi L}^{(1)}$ and
${\cal O}_{\Phi L}^{(3)}$ are redundant. \smallskip

As for the case of the bosonic operators, in order to obtain more
stringent bounds and to separate the contributions of the different
operators we consider specific initial states $X$. In particular
choosing the linear combinations as displayed in
Table~\ref{tab:fernogen} we are able to impose independent bounds in
each of the fermionic Wilson coefficients participating in the
$f\bar{f} \rightarrow VV$ scattering amplitudes. \smallskip

Starting from the states defined in Table~\ref{tab:fernogen} and the
scattering amplitudes given in Appendix~\ref{app:helamp} we obtain the
partial--wave helicity amplitudes also listed in
Table~\ref{tab:fernogen}.  The corresponding constraints on the Wilson
coefficients of the fermionic operators are
\begin{equation}
\begin{array}{l@{\hspace{2cm}}l@{\hspace{2cm}}l}
\left |\frac{f^{(1)}_{\Phi e} }{\Lambda^2} s\right| <53  \;\;\;,\;\;\;
& \left |\frac{f^{(1)}_{\Phi u} }{\Lambda^2}s\right | < 34 \;\;\;,\;\;\;
&\left |\frac{f^{(1)}_{\Phi d} }{\Lambda^2}s \right | < 23 \;\;\;,\;\;\;
\\
&&
\\
\left|\frac{f^{(1)}_{\Phi Q} }{\Lambda^2}s \right | < 14 \;\;\;,\;\;\;
&\left |\frac{f^{(3)}_{\Phi Q} }{\Lambda^2}s \right | < {123} \;\;\;,\;\;\;
&\left |\frac{f^{(1)}_{\Phi ud}}{\Lambda^2}s \right| < 13 \;\;\;.
\end{array} 
\label{eq:fernogenffbounds}
\end{equation}

\subsection{Bounds on Generation Dependent Operators}  

In this case first we need to eliminate the redundant operators
in Eq.~(\ref{eq:EOMred}). In order to do so we define four
independent combinations of the six leptonic operators
${\cal O}^{(3)}_{\Phi L,ii}$ and ${\cal O}^{(1)}_{\Phi L,ii}$ which are not removed by the EOM's. They are
\begin{equation}
\begin{array}{lll}
{\cal O}^{(1)}_{\Phi L,22-11}=  {\cal O}^{(1)}_{\Phi L,22}-{\cal O}^{(1)}_{\Phi L,11}\;  ,
&\;\;\;&
{\cal O}^{(1)}_{\Phi L,33-11}=  {\cal O}^{(1)}_{\Phi L,33}-{\cal
         O}^{(1)}_{\Phi L,11}\;  ,\\
&& \\
{\cal O}^{(3)}_{\Phi L,22-11}=  {\cal O}^{(3)}_{\Phi L,22}-{\cal O}^{(3)}_{\Phi L,11}\;  ,
&& 
{\cal O}^{(3)}_{\Phi L,33-11}=  {\cal O}^{(3)}_{\Phi L,33}-{\cal O}^{(3)}_{\Phi L,11}\;  ,\\
\end{array}
\label{eq:lepnouni}
\end{equation}
and we denote the corresponding Wilson coefficients as
$f^{(1)}_{\Phi L,22-11}$,
$f^{(1)}_{\Phi L,33-11}$,
$f^{(3)}_{\Phi L,22-11}$, and
$f^{(3)}_{\Phi L,33-11}$ respectively.  \smallskip

\begin{table}
\scalebox{0.8}{
\begin{tabular}{l|l}
  State & $T^{J=1}$  \\\hline
$\frac{1}{\sqrt{2 \Ng+{\Ng^2}}}
|{\displaystyle \sum_{i=1}^{\Ng}}
\left(e^-_{-,i}\,e^+_{+,i}\,+\,\nu_{-,i}\bar\nu_{+,i}\,\right)
{+}\,{\Ng}\,e^-_{+,j}\,e^+_{-,j} \rangle$
&
$ {-}i \frac{1}{12\sqrt{2}\pi} \frac{1}{\sqrt{2 \Ng+{\Ng^2}}} {\Ng}
\frac{f^{(1)}_{\Phi e,jj} \, s}{\Lambda^2}$
\\
\hline
$\frac{1}{\sqrt{8\Ng+{\Ng^2}\nc)}}
|{\displaystyle \sum_{i=1}^{\Ng}}
\left(2\,e^-_{-,i}\,e^+_{+,i}\,+2\,\,\nu_{-,i}\bar\nu_{+,i}\right)\, {-}
{\displaystyle\sum_{a=1}^{\nc}}{\Ng}\, u^a_{+,j}\,\bar u^a_{-,j}\rangle$
&
${+} i \frac{1}{12\sqrt{2}\pi} \frac{1}{\sqrt{8\Ng+{\Ng^2}\nc}}
{\Ng} \nc
\frac{f^{(1)}_{\Phi u,jj} \, s}{\Lambda^2}$
\\
\hline
$\frac{1}{\sqrt{2\Ng+{\Ng^2}\nc}}|{\displaystyle \sum_{i=1}^{\Ng}}
\left(\,e^-_{-,i}\,e^+_{+,i}\,+\,\,\nu_{-,i}\bar\nu_{+,i}\right)\, {+}
{\displaystyle\sum_{a=1}^{\nc}} {\Ng}\,d^a_{+,j}\,\bar d^a_{-,j}\rangle$
&
$ {-}i \frac{1}{12\sqrt{2}\pi} \frac{1}{\sqrt{2\Ng+ {\Ng^2}\nc}} {\Ng} \nc
\frac{f^{(1)}_{\Phi d,jj} \, s}{\Lambda^2}$
\\\hline
$\frac{1}{\sqrt{2\Ng+ {2 \Ng^2}\nc}}
|{\displaystyle \sum_{i=1}^{\Ng}}
\left(\,e^-_{-,i}\,e^+_{+,i}\,+\,\,\nu_{-,i}\bar\nu_{+,i}\right)\, +
{\displaystyle\sum_{a=1}^{\nc}} \Ng\, \left(d^a_{-,j}\,\bar d^a_{+,j}\,
+\, u^a_{-,j}\,\bar u^a_{+,j}\right) \rangle$
&
$ {+}i \frac{1}{12\sqrt{2}\pi} \frac{1}{\sqrt{2\Ng+{2\Ng^2}\nc}}
{\Ng} \nc\, 2\,
\frac{f^{(1)}_{\Phi Q,jj} \, s}{\Lambda^2}$
\\\hline
$\frac{1}{\sqrt{2\Ng+{2\Ng^2} }}
|{\displaystyle \sum_{i=1}^{\Ng}}
\left(\,e^-_{-,i}\,e^+_{+,i}\,-\,\,\nu_{-,i}\bar\nu_{+,i}\right)\, +
\,{\Ng} \,u^a_{-,j}\,\bar u^a_{+,j} \,-\, {\Ng}\,d^a_{-,j}\,\bar d^a_{+,j}\rangle$
&
${+} i \frac{1}{12\sqrt{2}\pi} \frac{1}{\sqrt{2\Ng+{2\Ng^2}}}
\frac{\Ng}{2}
\frac{f^{(3)}_{\Phi Q,jj} \, s}{\Lambda^2}$
\\\hline
$\frac{1}{\sqrt{\nc }}
|{\displaystyle \sum_{a=1}^{\nc}}
  d^a_{+,{j}}\,\bar u^a_{-,j} \rangle$
&  
$ -i \frac{1}{12\sqrt{2}\pi} \frac{1}{\sqrt{\nc}} \nc \sqrt{2}
\frac{f^{(1)}_{\Phi ud,jj} \, s}{\Lambda^2}$
\\\hline
$\frac{1}{{\sqrt{4}}}
|\,e^-_{-,1}\,e^+_{+,1}\,+ \,\nu_{-,1}\bar\nu_{+,1}
\,-\,e^-_{-,j}\,e^+_{+,j}\,- \,\nu_{-,j}\bar\nu_{+,j}\rangle $
&
${-}i \frac{1}{12\sqrt{2}\pi} {\frac{1}{2}2} 
\frac{f^{{(1)}}_{\Phi L,jj-11} \, s}{\Lambda^2}$
\\\hline
$\frac{1}{{\sqrt{4}}}
|\,e^-_{-,1}\,e^+_{+,1}\,- \,\nu_{-,1}\bar\nu_{+,1}
\,-\,e^-_{-,j}\,e^+_{+,j}\,+ \,\nu_{-,j}\bar\nu_{+,j}\rangle $
&
${+}i \frac{1}{12\sqrt{2}\pi} {\frac{1}{2}} \, {\frac{1}{2}}\,
\frac{f^{{(3)}}_{\Phi L,jj-11} \, s}{\Lambda^2}$
\\\hline
\end{tabular}  }
\caption{Initial fermionic states used to obtain independent bounds on the
  generation dependent fermionic operators. We also present the high
  energy   dominant term of the corresponding amplitudes. In the last two lines
  $j=2,3$. }
\label{tab:fergen}
\end{table}

It is interesting to notice that the sum over the three generations
for the $Q=0$ leptonic $+-00$ amplitudes cancel for the
left--handed operators because this is the combination removed by the
EOM.  With this in mind we define the initial states in
Table~\ref{tab:fergen} to impose bounds on each of the fermionic
operators. Using these initial states and the corresponding helicity
amplitudes, the bounds coming from $ f \bar{f} \to VV$ on each of the
Wilson coefficients of the fermionic operators read
\begin{equation}
\begin{array}{l@{\hspace{2cm}}l@{\hspace{2cm}}l}
\left |\frac{f^{(1)}_{\Phi e,jj} }{\Lambda^2}s\right |<69 \;\;\;,\;\;\;
& \left |\frac{f^{(1)}_{\Phi u,jj}}{\Lambda^2}s\right |<42 \;\;\;,\;\;\;
&\left |\frac{f^{(1)}_{\Phi d,jj} }{\Lambda^2}s\right |<34  \;\;\;,\;\;\;
\\
&&
\\
\left |\frac{f^{(1)}_{\Phi Q,jj} }{\Lambda^2}s\right |< {23} \;\;\;,\;\;\; 
& \left |\frac{f^{(3)}_{\Phi Q,jj} }{\Lambda^2}s \right |<{174} \;\;\;,\;\;\;
& \left |\frac{f^{(1)}_{\Phi ud,jj} }{\Lambda^2}s \right|<22  \;\;\;,\;\;\;
\\
&&
\\
\left |\frac{f^{{(1)}}_{\Phi L,jj-11} }{\Lambda^2}s\right |<{53} \;\;\;,\;\;\;
& \left |\frac{f^{{(3)}}_{\Phi L,jj-11} }{\Lambda^2}s\right |<{213}
\;\;\;,\;\;\; 
\end{array}
\label{eq:fergenffbounds}
\end{equation}
with $j=2,3$ in the last two inequalities.
As expected, the above limits are weaker than the ones displayed in
Eq.~(\ref{eq:fernogenffbounds}) for the generation independent
operators.  

\begin{table}[h!]\centering
\renewcommand{\arraystretch}{1.2}
\begin{tabular}{|c|c|c|c|}
\hline 
coupling & \multicolumn{3}{c|}{95\% allowed range  (TeV$^{-2})$} \\
\hline 
&  Generation Independent   & \multicolumn{2}{c|}{Generation Dependent}   \\
\hline\hline
$f_{BW}$     &  $(-0.32\,,\,1.7)$ 
&  \multicolumn{2}{c|}{$(-0.90\,,\,2.6)$ }
\\
$f_{\Phi 1}$ &  $(-0.040\,,\, 0.15)$
&  \multicolumn{2}{c|}
  {$(-0.11\,,\, 0.23)$}
\\
$f_{LLLL}$        & $(-0.043  \,0.013)$
& \multicolumn{2}{c|}{$(-1.3  \,,-0.21 )$}
\\
\hline
& & Case & Range \\\hline
$f^{(1)}_{\Phi Q}$
& $(-0.083  \,,\,0.10   )$
& $(11)=(22)$ & $(-0.33  \,,\,0.29)$
\\
$f^{(3)}_{\Phi Q}
$  & $(-0.60  \,,\, 0.12  )$
& $(11)=(22)$ & $(-0.92  \,,\, 0.64  )$
\\
& & $f^{(1)}_{\Phi Q,33}+\frac{1}{4}f^{(3)}_{\Phi Q,33}$ &
$(-0.21 \,,\, 0.041  )$
\\
\hline
$f^{(1)}_{\Phi u}$  & $(-0.25  \,,\, 0.37  )$
& $(11)=(22)$ &  $(-0.19  \,,\, 0.50  )$
\\
\hline
$f^{(1)}_{\Phi d}$  & $(-1.2  \,,\, -0.13  )$
& $(11)=(22)$  & $(-2.7  \,,\, 1.9  )$ \\
& & $(33)$  & $(-1.3  \,,\, -0.23  )$
\\
\hline
$f^{(1)}_{\Phi L}$
& ---- 
& $(22-11)$ & $(0.005  \,,\,0.41)$
\\
&
& $(33-11)$ & $(-0.63  \,,\,-0.096)$\\
\hline
$f^{(3)}_{\Phi L}$
& ---- 
& $(22-11)$ & $(-1.62  \,,\,-0.060)$
\\
&
& $(33-11)$ & $(0.38  \,,\,2.5)$\\
\hline
&  &  $(11)$      & $(-0.11  \,,\,0.049)$\\
$f^{(1)}_{\Phi e}$ & $(-0.075  \,,\,0.011  )$
&  $(22)$  & $(-0.15  \,,\,0.063)$\\
&  &  $(33)$  &  $(-0.15  \,,\,0.044)$\\
\hline
\end{tabular}
\caption{\it 95\% C.L. allowed ranges for the Wilson coefficients of the
  dimension--six operators that contribute to  the EWPD.}
\label{tab:ewpd}
\end{table}

\section{Discussion}
\label{sec:disc}

Let us start by noticing that even in the most general case, allowing for
all operators to have non--vanishing coefficients, we have obtained bounds
that are closed ranges. This means that there is a bounded region of the
parameter space for which  the effective theory is perturbatively valid.
In other words there is no combination of Wilson coefficients that can
extend indefinitely the energy domain where there is no partial--wave
unitarity violation.\smallskip

Second we want to address whether, within that region of coefficients,
violation of unitarity can be an issue at the Run~II LHC energies. Our
procedure to quantitatively answer this question is to determine the
maximum center--of--mass energy for which the unitarity limits are not
violated given our present knowledge on the Wilson coefficients of the
dimension--six operators from lower energy data.  For definiteness we
considered the maximum allowed value of these coefficients at the 95\%
confidence level in our analysis. Clearly the results depend
on this hypothesis and the energy range where perturbative unitarity
holds is extended if we consider these coefficients at 68\%
confidence level.\smallskip

In this respect EWPD gathered at the $Z$--pole and $W$--pole lead to
stringent constraints on operators contributing at linear order
to these observables and these results are model independent. These are
the fermionic operators leading
to $Z$ and $W$ couplings to fermions with the same Lorentz structure as the SM,
most of the fermionic operators in Eq.~\eqref{eq:hffop}, together  with
the bosonic operators ${\cal O}_{BW}$ and ${\cal O}_{\Phi,1}$.
The 95\% CL allowed range for their coefficients obtained from a global
analysis performed in the full multi--dimensional parameter space are
presented in Table \ref{tab:ewpd}; further details of the analysis
are given in Appendix~\ref{app:ewpd}. \smallskip

With these results we have quantified the maximum center--of--mass
energy for which partial--wave unitarity holds for each operator in
two scenarios. In the first we do not allow for
  cancellations among the contributions of the bosonic operators in
  the $s$-growing terms in $VV\rightarrow VV$ scattering, and
  therefore, we use the constraints obtained with just one
  non-vanishing Wilson coefficient; see Eq.~\eqref{eq:bound1vvvv}. In
  addition to that we considered generation independent fermion
  operators, Eq.~\eqref{eq:fernogenffbounds}, and the corresponding
  bounds on the Wilson coefficients in the central column in Table
  \ref{tab:ewpd}.  In the second scenario we use the unitarity
constraints on the bosonic operators allowing for cancellations in the
$VV\rightarrow VV$ scattering amplitudes, as in
Eq.~\eqref{eq:boundMvvvv}, together with the assumption of
generation--dependent fermion operators,
Eq.~\eqref{eq:fergenffbounds}, and the corresponding bounds on the
Wilson coefficients in the last column in Table \ref{tab:ewpd}.  The
maximum center--of--mass energy for which partial--wave unitarity
holds in both scenarios is:
\begin{equation}
\begin{array}{l@{\hspace{.5cm}}l@{\hspace{3cm}}l@{\hspace{.5cm}}l}
{\cal O}_{\Phi,1}   & \sqrt{s}_{\rm max} =18\,{\rm TeV}
\;\;\;,\;\;\;
&{\cal O}_{BW}   & \sqrt{s}_{\rm max} = 16.\,{\rm TeV}
\;\;\;,\;\;\;
\\
{\cal O}^{(1)}_{\Phi e}   & \sqrt{s}_{\rm max} =21\,{\rm TeV}
\;\;\;,\;\;\;
&{\cal O}^{(1)}_{\Phi u}   & \sqrt{s}_{\rm max} = 9.2 \,{\rm TeV} \;\;\;,\;\;\;
\\
  {\cal O}^{(1)}_{\Phi d}   & \sqrt{s}_{\rm max} =3.5\,{\rm TeV}
\;\;\;,\;\;\;
&{\cal O}^{(1)}_{\Phi Q}   & \sqrt{s}_{\rm max} = 8.3\,{\rm TeV} \;\;\;,\;\;\;
\\
{\cal O}^{(3)}_{\Phi Q}   & \sqrt{s}_{\rm max} =14\,{\rm TeV}
\;\;\;,\;\;\; & & \\
{\cal O}^{(1)}_{\Phi L,22-11}   & \sqrt{s}_{\rm max} =11\,{\rm TeV}
\;\;\;,\;\;\; &
{\cal O}^{(1)}_{\Phi L,33-11}   & \sqrt{s}_{\rm max} =9.2\,{\rm TeV}
\;\;\;,\;\;\;\\
{\cal O}^{(3)}_{\Phi L,22-11}   & \sqrt{s}_{\rm max} =12\,{\rm TeV}
\;\;\;,\;\;\; &
      {\cal O}^{(3)}_{\Phi L,33-11}   & \sqrt{s}_{\rm max}=9.2\,{\rm TeV}
\;\;\;.\;\;\;\\
\end{array}
\label{eq:smax1}
\end{equation}

Notice that the fermionic operator ${\cal O}^{(1)}_{\Phi ud}$ does not
contribute to the observables used in the $Z$--pole and $W$--pole data
analysis at the linear level as it gives a right--handed $W$ coupling
which does not interfere with the SM amplitude.  It does however,
contribute linearly to observables which depend on specific entries of
the CKM matrix via a finite renormalization of the quark mixing, in
particular to deep inelastic scattering of neutrinos off nucleons, as
well as, measurements of the CKM matrix elements in hadronic decays
~\cite{delAguila:2011zs,deBlas:2013gla}. The derivation of the bounds
on its coefficient from this data involves additional assumptions
about its flavour structure and the presence of further four--fermion
operators which also contribute to these processes, making them more
model dependent. Under the assumption of generation independent
couplings with no cancellation with the additional four--fermion
operators one obtains the constraints in
Refs.~\cite{delAguila:2011zs,deBlas:2013gla},
$-0.006\leq \frac{f^{(1)}_{\Phi ud}}{\Lambda^2} \leq 0.01$ which imply
$\sqrt{s}_{\rm max} =25.$ TeV. \smallskip

For the remaining dimension--six operators that we studied, the
present bounds on their Wilson coefficients come from global fits to
Higgs physics and TGC~\cite{Corbett:2015ksa,Butter:2016cvz} at the LHC
Run~I and the corresponding  maximum center--of--mass energy for which
partial--wave unitarity holds is:
\begin{equation}
\begin{array}{l@{\hspace{.5cm}}l@{\hspace{3cm}}l@{\hspace{.5cm}}l}
{\cal O}_{B}   & \sqrt{s}_{\rm max} =7.2\;{\rm TeV} \;\;\;,\;\;\;
&{\cal O}_{W}   & \sqrt{s}_{\rm max}=4.7\;{\rm TeV} \;\;\;,\;\;\;
\\
{\cal O}_{BB}   & \sqrt{s}_{\rm max} = 10.\;{\rm TeV} \;\;\;,\;\;\;
&{\cal O}_{WW}   & \sqrt{s}_{\rm max}= 5.2\;{\rm TeV} \;\;\;,\;\;\;
\\
{\cal O}_{\Phi,2}   & \sqrt{s}_{\rm max}=2.1\;{\rm TeV} \;\;\;,\;\;\;
&{\cal O}_{WWW}   & \sqrt{s}_{\rm max} = 5.7\;{\rm TeV} \;\;\;.\;\;\;
\end{array}
\label{eq:smax2}
\end{equation}

In order to access the importance of the above results for the
  LHC analyses we should keep in mind that, presently, the most energetic diboson  ($VV$) events possess a center--of--mass energy of the order of 3
  TeV; see for instance~\cite{CMS:2016djf}. As
  more integrated luminosity is accumulated this maximum energy will
  grow to 4--5 TeV, so we consider that as long as unitarity violation occurs
  only above these energies, it will not be an issue within the present LHC
  runs. This condition, of course, will have to be revisited at higher
  luminosity runs, but at that point also one will have to take into
  account the possible  more stringent bounds on the Wilson coefficients.
  \smallskip

  From the results in Eq.~\eqref{eq:smax1}--\eqref{eq:smax2}
  we read that there is no need of modification of
  the dimension--six effective theory to perform the LHC analyses for
  most operators. One exception is the operator
  ${\cal O}^{(1)}_{\Phi d}$ whose relatively lower
  $\sqrt{s}_{\rm max} = 3.5$ TeV, however, originates
  from the weaker bounds on its coefficients induced by the 2.8$\sigma$
  discrepancy of $A_{\rm FB}^{0,b}$ in the EWPD.
  Notwithstanding, studies of anomalous triple gauge couplings in
  diboson production should analyze more carefully the high energy
  tail of the distributions if they include this
  coupling. Furthermore, there is one additional exception that is the
  operator ${\cal O}_{\Phi,2}$. Since this operator modifies the
  production and decay of Higgs bosons, as well as, the $VV \to VV$
  scattering in vector boson fusion the high energy tails of these
  processes may also need a special scrutiny. \smallskip

An eventual caveat of the above conclusions is that the
UV completion might be strongly interacting and the lowest
center--of--mass energy exhibiting perturbative unitarity violation
then marks the onset of the strongly interacting region. If this were
the case at the LHC we should observe new states,
which has not yet been the case yet.


\section*{Acknowledgments}
We thank J. Gonzalez--Fraile for his valuable contributions to this
work. O.J.P.E. is supported in part by Conselho Nacional de Desenvolvimento
Cient\'{\i}fico e Tecnol\'ogico (CNPq) and by Funda\c{c}\~ao de Amparo
\`a Pesquisa do Estado de S\~ao Paulo (FAPESP); MCG-G 
is supported by USA-NSF grant PHY-1620628, by EU
Networks FP10 ITN ELUSIVES (H2020-MSCA-ITN-2015-674896) and
INVISIBLES-PLUS (H2020-MSCA-RISE-2015-690575), by MINECO grant
FPA2016-76005-C2-1-P and by Maria de Maetzu program grant MDM-2014-0367 of
ICCUB. T.C. is supported by the Australian Research Council.

\newpage

\appendix

\section {Helicity Amplitudes}
\label{app:helamp}

We present in this appendix the list of unitarity violating amplitudes
considered in this work that must be complemented by those in
Ref.~\cite{Corbett:2014ora}.

{\footnotesize
  \begin{table}[h!]
\tiny
  \begin{tabular}{|c||c|}
\hline
&$(\times \frac{f_{\Phi,1}}{\Lambda^2} \times s)$  
\\ \hline
$W^+W^+ \rightarrow W^+W^+$ &  $1$
\\ 
\hline
$W^+Z \rightarrow W^+Z$&$-\frac{X}{4}$
\\
$W^+ Z \rightarrow W^+ H$  &$\frac{2+Y}{4}$
\\
$W^+ H \rightarrow W^+ H$ &$-\frac{X}{4}$
\\ 
\hline
$W^+W^- \rightarrow W^+W^-$  &$-\frac{Y}{2} $
\\ 
$W^+W^- \rightarrow Z Z$ &  $\frac{1}{2}$
\\ 
$W^+W^- \rightarrow Z H$ &$\frac{Y-1}{2}$
\\
$W^+W^- \rightarrow H H$ &$-\frac{1}{2}$
\\ 
\hline
$Z Z \rightarrow  H H $ &$1$
\\
\hline
$Z H \rightarrow  Z H $ &$\frac{X}{2}$
\\   
\hline
\end{tabular}
\caption{Leading unitarity violating   terms of the scattering 
  amplitudes  for longitudinal gauge bosons generated by the operator
  ${\cal O}_{\Phi,1}$ where    $X=1-\cos\theta$ and $Y=1+\cos\theta$ 
and $\theta$ is the polar scattering angle.
}
\label{tab:Ophi124}
\end{table}}

\begin{table}[h]
\tiny
\begin{tabular}{|c|c|c|c|c|c|c|c|}
\hline
&\multicolumn{7 }{c|}{$(\times e^2 \frac{f_{BW}}{\Lambda^2})\times s$}   
\\
&$0000$&$00++$ & $0+0-$ & $0+-0$ & $+00-$ & $+0-0$ &
 $++00$ \\ \hline
$W^+W^+ \rightarrow W^+W^+$& 
0&$0$&$0$&$0$&$0$&$0$&$0$
\\ 
\hline
$W^+Z \rightarrow W^+Z$& 
0&$\frac{1}{4\cw}$&$\frac{1}{4}X$&
$\frac{1}{8\cw}Y$&$\frac{1}{8\cw}Y$&$0$&$\frac{1}{4\cw}$
\\
$W^+\gamma \rightarrow W^+\gamma$& 
$\na$&$\na$&$-\frac{1}{4}X$&$\na$&$\na$&$\na$&$\na$   
\\ 
$W^+Z \rightarrow W^+\gamma$& 
$\na$&$-\frac{1}{4\sw}$&$-\frac{1}{4}X\cot(2\theta_{\rm W})$&
$\na$&$-\frac{1}{8\sw}Y$&$\na$&$\na$   
\\ 
$W^+ Z \rightarrow W^+ H$& 
$0$&$\na$&$\na$&$\frac{1}{8\cw}Y$&$\na$&$\na$&
$-\frac{1}{4\cw}$   
\\ 
$W^+ \gamma \rightarrow W^+ H$ &
$\na$&$\na$&$\na$&$-\frac{1}{8\sw}Y$&$\na$&$\na$&
$\frac{1}{4\sw}$   
\\ 
$W^+ H \rightarrow W^+ H$ &
0&$\na$&$\na$&$\na$&$\na$&$0$&$\na$   
\\ 
 \hline
$W^+W^- \rightarrow W^+W^-$ &   
0&$0$&$0$&$0$&$0$&$0$&$0$   
\\ 
$W^+W^- \rightarrow Z Z$ &
0&$-\frac{1}{2}$&$-\frac{1}{8\cw}X$&$ {+}\frac{1}{8\cw}Y$&
${+}\frac{1}{8\cw}Y$&$-\frac{1}{8\cw}X$&$0$   
\\ 
$W^+W^- \rightarrow \gamma\gamma$ & 
$\na$&$\frac{1}{2}$&$\na$&$\na$&$\na$&$\na$&$\na$   
\\ 
$W^+W^- \rightarrow Z \gamma$ &
$\na$&$\frac{1}{2}\cot(2\theta_{\rm W})$&$\frac{1}{8\sw}X$&
$\na$&${-}\frac{1}{8\sw}Y$&$\na$&$\na$   
\\  
$W^+W^- \rightarrow Z H$ &
$0$&$\na$&$\na$&$\frac{1}{8\cw}Y$&$\na$&$
\frac{1}{8\cw}X$&$0$   
\\ 
$W^+W^- \rightarrow \gamma H$ & 
$\na$&$\na$&$\na$&$-\frac{1}{8\sw}Y$&$\na$&${-}\frac{1}{8\sw}X$&$\na$   
\\   
$W^+W^- \rightarrow H H$ &
0&$\na$&$\na$&$\na$&$\na$&$\na$&$0$   
\\ 
\hline
$Z Z \rightarrow  Z Z$ &
$0$&$\frac{1}{2}$&$-\frac{1}{4}X$&$\frac{1}{4}Y$&$\frac{1}{4}Y$
&$-\frac{1}{4}X$&$\frac{1}{2}$ 
\\ 
$Z Z \rightarrow  \gamma \gamma $ & 
$\na$&$-\frac{1}{2}$&$\na$&$\na$&$\na$&$\na$&$\na$ 
\\ 
$Z Z \rightarrow  Z \gamma $ & 
$\na$&$-\frac{1}{2}\cot(2\theta_{\rm W})$&
$\frac{1}{4}X\cot(2\theta_{\rm W})$&$\na$&
$-\frac{1}{4}Y\cot(2\theta_{\rm W})$&$\na$&$\na$ 
\\ 
$Z Z \rightarrow  Z H $ &
$0$&$\na$&$\na$&$0$&$\na$&$0$&$0$ 
\\ 
$Z Z \rightarrow  \gamma H $ & 
$\na$&$\na$&$\na$&$0$&$\na$&$0$&$0$ 
\\ 
$Z Z \rightarrow  H H $ &
0&$\na$&$\na$&$\na$&$\na$&$\na$&$-\frac{1}{2}$ 
\\ 
\hline
&$0000$&$00++$ & $0+0-$ & $0+-0$ & $+00-$ & $+0-0$ &
 $++00$ \\ \hline
\hline
$Z \gamma \rightarrow  Z Z$ &
$\na$&$\na$&$\frac{1}{4}X\cot(2\theta_{\rm W})$&
$-\frac{1}{4}Y\cot(2\theta_{\rm W})$&$\na$&$\na$&
$-\frac{1}{2}\cot(2\theta_{\rm W})$ 
\\ 
$Z \gamma \rightarrow  \gamma \gamma $ &
$\na$&$\na$&$\na$&$\na$&$\na$&$\na$&$\na$ 
\\ 
$Z \gamma \rightarrow  Z \gamma $ &
$\na$&$\na$&$\frac{1}{4}X$&$\na$&$\na$&$\na$&$\na$ 
\\ 
$Z \gamma \rightarrow  Z H $ &
$\na$&$\na$&$\na$&$0$&$\na$&$\na$&$0$ 
\\  
$Z \gamma \rightarrow  \gamma H $ & 
$\na$&$\na$&$\na$&$0$&$\na$&$\na$&$\na$ 
\\ 
$Z \gamma \rightarrow  H H $ &
$\na$&$\na$&$\na$&$\na$&$\na$&$\na$&$\frac{1}{2}\cot(2\theta_{\rm W})$ 
\\

\hline
$\gamma \gamma \rightarrow  \gamma \gamma $ &
$\na$&$\na$&$\na$&$\na$&$\na$&$\na$&$\na$ 
\\ 
$\gamma \gamma \rightarrow  H H $ &
$\na$&$\na$&$\na$&$\na$&$\na$&$\na$&$\frac{1}{2}$ 
\\ 
$Z H \rightarrow  Z H $ &
0&$\na$&$\na$&$\na$&$\na$&$-\frac{1}{4}X$&$\na$    
\\  
$Z H \rightarrow  Z \gamma $ &
$\na$&$0$&$\na$&$\na$&$0$&$\na$&$\na$ 
\\ 
$\gamma H \rightarrow  \gamma H $ &
$\na$&$\na$&$\na$&$\na$&$\na$&$\frac{1}{4}X$&$\na$ 
\\ 
\hline
$Z H \rightarrow \gamma H$ &
$\na$&$\na$&$\na$&$\na$&$\na$&
$\frac{1}{4}X\cot(2\theta_{\rm W})$&$\na$  \\
\hline
\end{tabular}
\caption{Leading unitarity violating terms of the scattering 
amplitudes  for  gauge bosons with the different helicities 
generated by the operator ${\cal O}_{BW}$.  
$X=1-\cos\theta$ and $Y=1+\cos\theta$ and $\theta$ is the polar
scattering angle.}
\label{tab:4Vother}
\end{table}

\begin{table}[h]
\scalebox{0.75}{
  \begin{tabular}{|l|c|l|l|}
\hline 

Process  & $\sigma_1,\sigma_2,\lambda_3,\lambda_4$   &  Bosonic
                                                       operator contribution
&  fermionic operator contribution \\
&    & $\left(\times\,  i\, \frac{s}{\Lambda^2}\times \sin\theta\right)$&
$\left(\times\, i\, \frac{s}{\Lambda^2}\times \sin\theta\right)$   
\\ \hline
$e^-_i e^+_i \to W^+W^-$
&	$-+00$	&
$-\frac{ g^2 }{8}\frac{\cwsq\fw+\swsq\fb}{\cwsq}	$
& $\frac{1}{4}(f^{(3)}_{\phi L, ii}-4 f^{(1)}_{\phi L, ii}) $
\\
&	$+-00$	&
$-\frac{ g^2}{4}\frac{\swsq\fb}{\cwsq}		$
&$ -  f^{(1)}_{\phi e, ii} $
\\
&	$-+- -$	&
$-\frac{3 g^4 }{8} f_{WWW}$ & $0$
\\
&	$-+++$	&
$-\frac{3 g^4}{8} f_{WWW}$
& $0$ \\
\hline 
$\nu_i \bar{\nu}_i \to W^+W^-$
&	$-+00$	&
$\frac{ g^2 }{8}\frac{\cwsq\fw-\swsq\fb}{\cwsq}$
& $-\frac{1}{4} (f^{(3)}_{\phi L, ii}+4 f^{(1)}_{\phi L, ii})$
\\
&	$+-00$	&$0$ & $0$\\
&	$-+- -$	&
$\frac{3 g^4 }{8} f_{WWW}$
& $0$
\\
&	$-+++$	&
$\frac{3 g^4}{8} f_{WWW}$
& $0$
\\
\hline
$u_i \bar{u}_i \to W^+W^-$
&	$-+00$	&
$\frac{ g^2 \nc}{8}\frac{3\cwsq\fw+\swsq\fb}{3\cwsq}$
& $-\frac{ \nc\,}{4}
(f^{(3)}_{\phi Q, ii}+4 f^{(1)}_{\phi Q, ii}) $
\\
&	$+-00$	&
$\frac{ g^2 \nc}{6}\frac{\swsq}{\cwsq}\fb$
& $  - \nc\,  f^{(1)}_{\phi u, ii}$
\\
&	$-+- -$	&
$\frac{3 g^4\nc}{8}\, f_{WWW} $
& $0$
\\
&	$-+++$	&
$\frac{3 g^4\nc}{8}\, f_{WWW}$
& $0$\\
\hline
$d_i \bar{d_i}\to W^+W^-$
&	$-+00$	&
$-\frac{  g^2 \nc }{8}\frac{3\cwsq\fw-\swsq\fb}{3\cwsq}$
& $\frac{ \nc\, }{4} (f^{(3)}_{\phi Q, ii}-4 f^{(1)}_{\phi Q, ii}) $
\\
&	$+-00$	&
$-\frac{  g^2 \nc }{12}\frac{\swsq\fb}{\cwsq}$
& $-  \nc\,  f^{(1)}_{\phi d, ii}$
\\
&	$-+- -$	&
$-\frac{3 g^4\nc }{8}\, f_{WWW}$
& $0$
\\
&	$-+++$	&
$-\frac{3 g^4\nc}{8}\,f_{WWW}$
& $0$\\
\hline
$e^-_i \bar{\nu_i}\to W^-Z$
&	$-+00$	&
$\frac{  g^2}{4\sqrt{2}}\,\fw $
& $-\frac{1}{2\sqrt{2}}\,f^{(3)}_{\phi L, ii}$
\\
&	$+-00$	&$0$ &$0$\\
&	$-+- -$	&
$\frac{3\cw g^4 }{4\sqrt{2}}\,f_{WWW}$
&$0$\\
&	$-+++$	&
$\frac{3 \cw g^4}{4\sqrt{2}}\, f_{WWW}$
&$0$\\
\hline
$e^-_i\bar{\nu_i}\to W^-A$
&	$-+00$	&$0$ & $0$\\
&	$+-00$	&$0$ & $0$\\
&	$-+- -$	&
$\frac{3 \sw g^4  }{4\sqrt{2}}\,f_{WWW}$
& $0$\\
&	$-+++$	&
$\frac{3 \sw g^4 }{4\sqrt{2}}\,f_{WWW}$
& $0$\\
\hline
$d_i \bar{u_i}\to W^-Z$
&	$-+00$	&
$\frac{  g^2 \nc }{4\sqrt{2}}\,\fw$
& $-\frac{\nc}{2\sqrt{2}}\,f^{(3)}_{\phi Q, ii}$ 
\\
&	$+-00$	&$0$
& $- \nc\sqrt{2}\, f^{(1)}_{\phi ud, ii}$
\\
&	$-+- -$	&
$\frac{3 \cw g^4\nc }{4\sqrt{2}}\,f_{WWW}$
&$0$
\\
&	$-+++$	&
$\frac{3 \cw g^4\nc}{4\sqrt{2}}\,f_{WWW}$
&$0$
\\
\hline
$ d_i  \bar{u_i}\to W^-A$
&	$-+00$	&$0$ &$0$\\
&	$+-00$	&$0$&$0$\\
&	$-+- -$	&
$\frac{3 \sw g^4 \nc }{4\sqrt{2}}\,f_{WWW}$
&$0$\\
&	$-+++$	&
$\frac{3 \sw g^4 \nc }{4\sqrt{2}}\,f_{WWW}$
&$0$\\
\hline 					
\end{tabular}
}
\caption{Leading unitarity violating terms of the scattering
  amplitudes $\mathcal{M}({f_1}_{\sigma_1}{\bar{f_2} }_{\sigma_2} \to
  {V_3}_{\lambda_3}{V_4}_{\lambda_4})$.
  Notice that in writing these amplitudes we have not
  imposed the conditions in Eq.~(\ref{eq:EOMred}) yet. See the text for details. }
\label{tab:fermiontable}
\end{table}

\section{Constraints from EWPD}
\label{app:ewpd}

We briefly summarize here the details of our analysis of EWPD.
Similar analyses for different choices of operator basis can be found
in~\cite{Pomarol:2013zra, Ciuchini:2014dea, Falkowski:2014tna,
  Efrati:2015eaa, Ellis:2014jta}. We work on the Z--scheme where the
input parameters are chosen to be
$\alpha_s, G_F, \alpha_\text{em}, M_Z$ ~\cite{Olive:2016xmw}, and
$M_h$ ~\cite{Aad:2015zhl}.  In addition to these quantities we also
consider the fermion masses as input parameters. All the other
quantities appearing in the Lagrangian are implicitly expressed as
combinations of experimental inputs. \smallskip

In our analyses we evaluated the dimension--six contributions to the
observables keeping both SM contribution and the linear
terms in the anomalous couplings, {\em i.e.} the we considered only
the interference between the SM and the anomalous contributions.  The
predictions for the shift in the observables of the $Z$ and $W$ pole
physics with respect to their SM values are
\begin{eqnarray}
&&\Delta \Gamma_Z=2 \Gamma_{Z,\rm SM} \left(
\frac{\displaystyle{\sum_f} (g_L^f\Delta g_L^f+g_R^f\Delta g_R^f)N_C^f}{
\displaystyle{\sum_f}(|g_L^f|^2+|g_R^f|^2)N_C^f}\right) \;\;,
\\
&&\Delta\sigma_h^0=
2\sigma_{h,\rm SM}^0\left(
\frac{(g_L^e\Delta g_L^e+g_R^e\Delta g_R^e)}{|g_L^e|^2+|g_R^e|^2}
+\frac{\displaystyle{\sum_q} (g_L^q\Delta g_L^q+g_R^q\Delta g_R^q)}
{\displaystyle{\sum_q}(|g_L^q|^2+|g_R^q|^2)}
-\frac{\Delta \Gamma_Z}
{\Gamma_{Z,\rm SM}}\right) \;\;,
\\
&&\Delta R_l^0\equiv
\Delta\left(\frac{\Gamma_Z^{\rm had}}{\Gamma_Z^{l}}\right)=
2 R_{l,\rm SM}^0
\left(\frac{\displaystyle{\sum_q} (g_L^q\Delta g_L^q+g_R^q\Delta g_R^q)}
{\displaystyle{\sum_q}(|g_L^q|^2+|g_R^q|^2)}
- \frac{(g_L^l\Delta g_L^l+g_R^l\Delta g_R^l)}{|g_L^l|^2+|g_R^l|^2}
\right) \;\;,
\\
&&\Delta R_q^0\equiv\Delta\left(\frac{\Gamma_Z^{q}}{\Gamma_Z^{\rm had}}\right)=
2R_{q,\rm SM}^0\left(
\frac{(g_L^q\Delta g_L^q+g_R^q\Delta g_R^q)}
{|g_L^q|^2+|g_R^q|^2}
- \frac{\displaystyle{\sum_{q'}} (g_L^{q'}\Delta g_L^{q'}+
g_R^{q'}\Delta g_R^{q'})}
{\displaystyle{\sum_{q'}}(|g_L^{q'}|^2+|g_R^{q'}|^2)}
\right) \;\;,
\end{eqnarray}
\begin{eqnarray}
&&\Delta\mathcal{A}_f=4 \mathcal A_{f,\rm SM}\;
\frac{g_L^fg_R^f}{|g_L^f|^4-|g_R^f|^4}
\left(g_R^f\Delta g_L^f-g_L^f\Delta g_R^f\right) \;\;,
\\
&&\Delta P_\tau^{\rm pol}=\Delta\mathcal{A}_l \;\;,
\\
&&\Delta A_{\rm  FB}^{0,f}=A_{\rm  FB,SM}^{0,f}\left(
\frac{\Delta\mathcal{A}_l}{\mathcal{A}_l}+
\frac{\Delta\mathcal{A}_f}{\mathcal{A}_f}\right) \;\;,
\\
&&\Delta \Gamma_W=\Gamma_{W,\rm SM}
\left(\frac{4}{3}\Delta g_{WL}^{ud}+\frac{2}{3}\Delta
   g_{WL}^{e\nu}+\Delta M_W\right) \;\;,
\\
&&\Delta Br_W^{e\nu}=Br_{W,\rm SM}^{e\nu}
\left(-\frac{4}{3}\Delta g_{WL}^{ud}+\frac{4}{3}\Delta g_{WL}^{e\nu}\right)\,,
\end{eqnarray}
where we write the induced corrections to
the SM fermion couplings of the Z boson ($g^f_{L(R)}$) as
\begin{equation}
\Delta g_{L,R}^f=g_{L,R}^f\Delta g_1+Q^f\Delta g_{2}+\Delta\tilde g^f_{L,R}\,.
\label{eq:deltags}
\end{equation}
The universal shifts of the fermion couplings in Eq.~(\ref{eq:deltags})
due to dimension--six operators are
\begin{eqnarray}
\Delta g_1=\frac{1}{2}\left(\alpha\, T-\frac{\delta G_F}{G_F}\right) \,,
&&
\Delta g_2=\frac{\st^2}{\cdt}\left(\ct^2 \left(\alpha\, T-\frac{\delta G_F}{G_F}\right) -
\frac{1}{4\st^2}\alpha\, S\right)\,.
\label{eq:dg12}
\end{eqnarray}
where we denoted the sine (cosine) of the weak mixing angle by
$\st$ ($\ct$). The cosine and sine of twice $\theta_W$ are then denoted
$\cdt$ and $\sdt$ respectively.
The tree level contributions of the dimension--six operators to the
oblique parameters~\cite{Peskin:1990zt, Peskin:1991sw} are
\begin{equation}
\begin{array}{llll}
  \alpha\,S=-e^2\frac{v^2}{\Lambda^2}\fbw \,,
  &\alpha\, T= -\frac{1}{2}\frac{v^2}{\Lambda^2}\fpone \, ,  & 
\alpha\, U=0\, , &
\frac{\delta G_F}{G_F}=-2
f_{LLLL}\frac{v^2}{\Lambda^2} + (f^{(3)}_{\Phi L,11}  +f^{(3)}_{\Phi
                   L,22})  {\frac{vˆ2}{\Lambda^2}}
\end{array}
\end{equation}
where for completeness we also included the effect of the
dimension--six four--fermion operator contributing with a finite
renormalization to the Fermi constant
\begin{equation}
  {\cal O}_{LLLL}=(\bar L \gamma^\mu L)(\bar L \gamma^\mu L) \;.
\end{equation}

The coupling modifications that depend on the
fermion flavor are given by
\begin{equation}
\begin{array}{lll}
  \Delta \tilde g^u_{L}=
-\frac{v^2}{8 \Lambda^2} (4f^{(1)}_{\Phi Q}- f^{(3)}_{\Phi Q}) \, , 
\qquad\qquad
&&
\Delta \tilde g^u_{R}= -\frac{v^2}{2 \Lambda^2}  f^{(1)}_{\Phi u} \, , 
\\
\Delta \tilde g^d_{L}= -\frac{v^2}{8 \Lambda^2}
(4f^{(1)}_{\Phi Q}+ f^{(3)}_{\Phi Q}) \, ,
&&
\Delta \tilde g^d_{R}= -\frac{v^2}{2 \Lambda^2}  f^{(1)}_{\Phi d} \,, \\
\Delta \tilde g^\nu_{L}=-\frac{v^2}{8 \Lambda^2} (4f^{(1)}_{\Phi L}- f^{(3)}_{\Phi L}) \,  ,
&&
\Delta \tilde g^\nu_{R}= 0\,,\\
\Delta \tilde g^e_{L}=-\frac{v^2}{8 \Lambda^2} (4f^{(1)}_{\Phi L}+ f^{(3)}_{\Phi L}) \,  \,,
&&
\Delta \tilde g^e_{R}=-\frac{v^2}{2 \Lambda^2}  f^{(1)}_{\Phi e}
\,.
\end{array}
\label{eq:coupmod}
\end{equation}

As for the couplings of the $W$ to fermions, 
in the SM we normalize the left (right)--handed couplings 
as 1 (0)  and the corresponding shifts on these couplings due
dimension--six operators are
\begin{equation}
  \Delta g_{WL}^{ff'}=\Delta g_W+ \Delta\tilde g_{WL}^{ff'}\, ,
  \;\;\;\;\;\;\;\;\;
  \Delta g_{WR}^{ff'}=\Delta\tilde g_{WR}^{ff'}\, ,
\label{eq:deltagw}
\end{equation}
with the universal shift given by
\begin{equation}
\Delta g_W=\frac{\Delta M_W}{M_W}-\frac{1}{2}\frac{\delta G_F}{G_F}\,, 
\end{equation}
where the correction to the $W$ mass coming from the dimension--six
operators reads
\begin{equation}
\frac{\Delta M_W}{M_W}=\frac{\ct^2}{2\cdt} \alpha\, T -\frac{1}{4\cdt} \alpha\, S
+\frac{1}{8\st^2}\alpha\, U -\frac{\st^2}{2\cdt} \frac{\delta G_F}{G_F}\; . 
\label{eq:dmw}
\end{equation}
The fermion dependent contributions of the dimension--six
operators to the $W$-couplings are
\begin{eqnarray}
  \Delta\tilde g_{WL}^{ud}=\frac{v^2}{4 \Lambda^2}  f^{(3)}_{\Phi Q} \,,
    &\;\;\;
    \Delta\tilde g_{WR}^{ud}=\frac{v^2}{\Lambda^2}  f^{(1)}_{\Phi ud} \,,
    & \;\;\; {\Delta\tilde g_{WL}^{e\nu}}=\frac{v^2}{4 \Lambda^2}  f^{(3)}_{\Phi L} \,,
\;\;\;\;\;
    \Delta\tilde g_{WR}^{e\nu}=0
     \,. 
\label{eq:dgw}
\end{eqnarray}

We notice that, as we are including the effect of the operators in the
observables at linear order, the operator ${\cal O}^{(1)}_{\Phi ud,ij}$
does not contribute since it leads to a right-handed CC current which
does not interfere with the corresponding SM amplitude. \smallskip

We perform two different fits which differ on the assumptions on the
generation dependence of the fermionic operators.

\subsection{Fit with  Generation Independent Operators}

In the first case we assume that the fermionic operators are
generation independent. In this case, as discussed above, we can drop
the generation index in all coefficients. Furthermore removing the
operators in Eq.~(\ref{eq:EOMred}) implies that those two operators do
not appear in the fit to the EWPD. We have then 8 coefficients to be
determined
\begin{equation*}
\left \{
\frac{f_{BW}}{\Lambda^2} \;\;,\;\;
\frac{f_{\Phi, 1}}{\Lambda^2} \;\;,\;\;
\frac{f_{LLLL}}{\Lambda^2} \;\;,\;\;
\frac{f^{(1)}_{\Phi Q}}{\Lambda^2} \;\;,\;\;
\frac{f^{(3)}_{\Phi Q}}{\Lambda^2} \;\;,\;\;
\frac{f^{(1)}_{\Phi u}}{\Lambda^2} \;\;,\;\;
\frac{f^{(1)}_{\Phi d}}{\Lambda^2} \;\;,\;\;
\frac{ f^{(1)}_{\Phi e}}{\Lambda^2}  \;\;
\right \} \;\;.
\end{equation*}
In our analyses we fitted 15 observables of which 12 are $Z$
observables~\cite{ALEPH:2005ab}:
\begin{equation*}
\Gamma_Z \;\;,\;\;
\sigma_{h}^{0} \;\;,\;\;
{\cal A}_{\ell}(\tau^{\rm pol}) \;\;,\;\;
R^0_\ell \;\;,\;\;
{\cal A}_{\ell}({\rm SLD}) \;\;,\;\;
A_{\rm FB}^{0,l} \;\;,\;\;
R^0_c \;\;,\;\;
 R^0_b \;\;,\;\;
{\cal  A}_{c} \;\;,\;\;
 {\cal A}_{b} \;\;,\;\;
A_{\rm FB}^{0,c}\;\;,\;\;
\hbox{ and} \;\;
A_{\rm FB}^{0,b}  \hbox{ (SLD/LEP-I)}\;\;\; ,
\end{equation*}
supplemented by three $W$ observables
\begin{equation*}
  M_W   \;\;,\;\; \Gamma_W \;\;,\hbox{  and}\;\; \hbox{Br}( W\to {\ell\nu})
\end{equation*}
that are, respectively, its average mass from~\cite{Olive:2016xmw},
its width from LEP-II/Tevatron~\cite{ALEPH:2010aa}, and the leptonic
$W$ branching ratio for which the average in Ref.~\cite{Olive:2016xmw}
is taken.  The correlations among the inputs can be found in
Ref.~\cite{ALEPH:2005ab} and have been taken into consideration in the
analyses. The SM prediction and its uncertainty due to variations
of the SM parameters are taken from~\cite{Ciuchini:2014dea}.
\smallskip

When performing the fit within the context of the SM the result is
$\chi^2_{\rm EWPD,SM}=18.0$, while including
the 8 new parameters it gets reduced to $\chi^2_{\rm EWPD,min}=5.3$.
The results of the analysis are shown in Table~\ref{tab:ewpd} where we
quote the 95\% C.L. allowed ranges for each parameter in the center
column. The range for parameter $x$ is obtained accounting for all
possible cancellations in the multiparameter space by imposing the
condition $\Delta\chi^2_{\rm EWPD, marg}(x) <4$ where by
$\Delta\chi^2_{\rm EWPD, marg}(x)$ we denote the value of
$\Delta\chi^2_{\rm EWPD}$ minimized with respect to the other seven
parameters for each value of the parameter $x$.  We notice that the only
operator coefficient not compatible with zero at $2\sigma$ is
$f^{(1)}_{{\Phi d}}$, a result driven by the 2.7$\sigma$
discrepancy between the observed $A_{\rm FB}^{0,b}$ and the SM
expectation.  \smallskip

\subsection{Fit with Generation Dependent Operators}

Lifting the assumption of generation independent operators we are left
with seven independent leptonic operators. These are, three
${\cal O}^{(1)}_{\Phi e,ii}$ plus four combinations of
${\cal O}^{(1)}_{\Phi L,ii}$ and ${\cal O}^{(3)}_{\Phi L,ii}$ defined
in Eq.~(\ref{eq:lepnouni}). On the other hand, for operators involving
quarks there is not enough information in the observables considered to
resolve the contributions from the two first generations. Consequently 
we make the simplifying assumption that operators for the first
and second generations have the same Wilson coefficients and only
those from the third generation are allowed to be different.
Furthermore, for the third generation of quarks only
${\cal O}^{(1)}_{\Phi Q,33}$ and the linear combination
$f^{(1)}_{\Phi Q,33} +\frac{1}{4} f^{(3)}_{\Phi Q,33}$ contribute independently
to  the $Z$ and $W$ observables; see Eq.~(\ref{eq:coupmod}). 
Altogether there are  a total of 16 coefficients to be determined from
the fit to the $Z$ and $W$ observables:
\begin{equation}  
\begin{array}{l@{\hspace{.5cm}}l@{\hspace{.5cm}}l@{\hspace{.5cm}}l}
\frac{f_{BW} }{\Lambda^2} \;\;\; ,
& \frac{ f_{\Phi 1} }{\Lambda^2}\;\;\;,
& \frac{ f_{LLLL} }{\Lambda^2}\;\;\;,
& \frac{ f^{(1)}_{\Phi Q,11}}{\Lambda^2}=
\frac{ f^{(1)}_{\Phi Q,22} }{\Lambda^2}\;\;\;,
\\
&&&\\
\frac{ f^{(1)}_{\Phi Q,33} }{\Lambda^2}{+}\frac{1}{4} 
\frac{ f^{(3)}_{\Phi Q,33} }{\Lambda^2}\;\;\;,
& \frac{ f^{(3)}_{\Phi Q,11 } }{\Lambda^2}=
\frac{ f^{(3)}_{\Phi Q,22} }{\Lambda^2}\;\;\;, 
& \frac{f^{(1)}_{\Phi u,11}}{\Lambda^2}=
\frac{f^{(1)}_{\Phi u,22} }{\Lambda^2}\;\;\;,
& \frac {f^{(1)}_{\Phi d,11} }{\Lambda^2}=
\frac{f^{(1)}_{\Phi d,22} }{\Lambda^2}\;\;\;,
\\
&&&\\
\frac{ f^{(1)}_{\Phi d,33} }{\Lambda^2}\;\;\;, 
& \frac{ f^{(1)}_{\Phi e,11} }{\Lambda^2}\;\;\;,
& \frac{ f^{(1)}_{\Phi e,22} }{\Lambda^2}\;\;\;,
& \frac{ f^{(1)}_{\Phi e,33} }{\Lambda^2}\;\;\;,
\\
&&&\\
\frac{ f^{(1)}_{\Phi L,22-11} }{\Lambda^2}\;\;\;,
&\frac{ f^{(1)}_{\Phi L,33-11} }{\Lambda^2}\;\;\;,
&\frac{ f^{(3)}_{\Phi L,22-11}}{\Lambda^2} \;\;\;,
& \frac{ f^{(3)}_{\Phi L,33-11} }{\Lambda^2} \;\;\;.
\end{array}
\end{equation}
\medskip

In order to obtain the corresponding constraints on these 16 parameters
a fit including 24 experimental data points is performed. These are 
19  $Z$ observables ~\cite{ALEPH:2005ab}: 
\[
\begin{array}{l@{\hspace{.25cm}}l@{\hspace{.25cm}}l@{\hspace{.25cm}}l@{\hspace{.25cm}}l}
\Gamma_Z \;\;, 
&\sigma_{h}^{0} \;\;, 
& R^0_e \;\;,
&R^0_\mu \;\;,
& R^0_\tau \;\;,
\\
A_{\rm  FB}^{0,e} \;\;,
& A_{\rm  FB}^{0,\mu} \;\;,
& A_{\rm  FB}^{0,\tau} \;\;,
& \mathcal{A}_{e}({\rm }\tau^{\rm pol}) \;\;, 
&\mathcal{A}_{\mu}({\rm } \tau^{\rm pol}) \;\;, 
\\
\mathcal{A}_{e}({\rm SLD}) \;\;,
&\mathcal{A}_{\mu}({\rm SLD}) \;\;,
& \mathcal{A}_{\tau}({\rm SLD}) \;\;, 
& R^0_c \;\;, 
&R^0_b \;\;, 
\\
\mathcal{A}_{c} \;\;,
& \mathcal{A}_{b} \;\;, 
&A_{\rm  FB}^{0,c} \;\;, \hbox{ and}
& A_{\rm  FB}^{0,b} \;\;,
\end{array}
\]
plus five $W$ observables: 
\[
    M_W \;\;\;\;,\;\;\;\;
    \Gamma_W \;\;\;\;,\;\;\;
     \hbox{Br} (W \to {e\nu}) \;\;\;\;,\;\;\;\;
      \hbox{Br}(W \to {\mu\nu}) \;\;\;\;, \hbox{ and }
      \hbox{Br}(W \to {\tau\nu}) \;\;\;\;,\;\;\;\;
\]
where the three leptonic $W$ branching ratios were taken from
Ref.~\cite{Olive:2016xmw}. The correlations among the inputs can be
found in Refs.~\cite{ALEPH:2005ab,Olive:2016xmw} and were considered
in the analysis. As in the previous analysis,  the SM  prediction for
these observables and its uncertainty due to variations
of the SM parameters are taken from~\cite{Ciuchini:2014dea}.
\smallskip

The fit within the context of the SM leads to $\chi^2_{\rm
  EWPD,SM}=29$, while with the inclusion of the 16 new parameters the minimum gets
reduced to $\chi^2_{\rm EWPD,min}=8.2$. The 95\% allowed ranges for
each of the 16 parameters are shown in the last column in
Table~\ref{tab:ewpd}. As in the previous case, for each coupling the
range is obtained after marginalization over the other 15 couplings.
\smallskip

As we can see from Table~\ref{tab:ewpd}, removing the generation
independence hypothesis leads to looser constraints, as could be
anticipated. Moreover, flavor independent Wilson coefficients and the
ones related to the first two families agree with the SM at the 2$\sigma$
level, with the exception of $f^{(1)}_{\Phi L,22-11}/\Lambda^2$. On the other
hand, we can see clearly the effect of the observable $A_{\rm FB}^{0,b}$
on almost all the  third generation Wilson coefficients whose $2\sigma$ allowed
ranges do not contain the SM. \smallskip

\section{Dipole operators}
\label{app:dipole}

The leading high energy contributions of the dipole fermionic
operators in Eq.~(\ref{eq:dipole}) to the $f \bar{f} \to VV$
scattering is given in Table~\ref{tab:DipoleAmp}. Neglecting fermion
masses the dipole
fermionic operators contribute to different helicity states to
those from operators in Eq.~\eqref{eq:hffop} as can be seen from
Tables~\ref{tab:fermiontable} and~\ref{tab:DipoleAmp}, due to the
presence of $\sigma^{\mu\nu}$ in Eq.~(\ref{eq:dipole}).\smallskip

\begin{table}[h]
\scalebox{0.75}{
  \begin{tabular}{|l|c|l|}
\hline 
Process  & $\sigma_1,\sigma_2,\lambda_3,\lambda_4$   &  partial-wave amplitude
\\& &  $\left(\times\, \frac{s}{\Lambda^2}\times \sin\theta\right)$
\\ \hline
$e^-_i e^+_i \to W^+W^-$
& $--0+$ & $-g\, f_{eW,ii} $ \\
& $++-0$ & $g\, f_{eW,ii} $ \\
\hline
$e^-_i e^+_i \to Z  Z $ 
& $--0+$ & $-\frac{g}{2 \cw}\, (f_{eW,ii} \cwsq + f_{eB,ii}\swsq)$\\
& $--+0$ & $-\frac{g}{2 \cw}\,  (f_{eW,ii} \cwsq + f_{eB,ii}\swsq)$\\
& $++0-$ & $\frac{g}{2 \cw}\,   (f_{eW,ii} \cwsq + f_{eB,ii}\swsq)$\\
& $++-0$ & $\frac{g}{2 \cw}\,   (f_{eW,ii} \cwsq + f_{eB,ii}\swsq)$\\
\hline
$u_i \bar{u}_i \to W^+W^-$
& $--+0$ & $-g\,   \nc \,f_{uW,ii} $ \\
& $++0-$ & $g\,   \nc\, f_{uW,ii} $ \\
\hline
$u_i \bar u_i \to Z  Z $ 
& $--0+$ & $-\frac{g}{2 \cw}\, \nc\, (f_{uW,ii} \cwsq - f_{uB,ii}\swsq)$\\
& $--+0$ & $-\frac{g}{2 \cw}\,  \nc\,(f_{uW,ii} \cwsq - f_{uB,ii}\swsq)$\\
& $++0-$ & $\frac{g}{2 \cw}\,   \nc\, (f_{{uW},ii} \cwsq - f_{{uB},ii}\swsq)$\\
& $++-0$ & $\frac{g}{2 \cw}\,   \nc\, (f_{{uW},ii} \cwsq - f_{{uB},ii}\swsq)$\\
\hline 
$d_i \bar{d_i}\to W^+W^-$
& $--0+$ & $-g\,\nc\,f_{dW,ii} $ \\
& $++-0$ & $g\,\nc\, f_{dW,ii} $ \\
\hline
$d_i \bar d_i \to Z  Z $ 
& $--0+$ & $-\frac{g}{2 \cw}\, \nc\,(f_{dW,ii} \cwsq + f_{dB,ii}\swsq)$\\
& $--+0$ & $-\frac{g}{2 \cw}\,  \nc\,(f_{dW,ii} \cwsq + f_{dB,ii}\swsq)$\\
& $++0-$ & $\frac{g}{2 \cw}\,  \nc\,(f_{dW,ii} \cwsq + f_{dB,ii}\swsq)$\\
& $++-0$ & $\frac{g}{2 \cw}\,   \nc\,(f_{dW,ii} \cwsq + f_{dB,ii}\swsq)$\\
\hline
$e^-_i \bar{\nu_i}\to W^-Z$
& $++-0$ & $-\frac{ g }{\sqrt{2}} f_{eW,ii}$\\
& $++0-$ & $-\frac{ g }{\sqrt{2}\cw}(f_{eB,ii} \swsq -f_{eW,ii} \cwsq)$\\
\hline
$e^-_i\bar{\nu_i}\to W^-A$
& $++0-$ & $\frac{ g\, \sw  }{\sqrt{2}}(f_{eB,ii}+f_{eW,ii})$
\\
\hline
$d_i \bar{u_i}\to W^-Z$
& $--0+$ & $\frac{ g\, \nc}{\sqrt{2}} (f_{uB,ii} \swsq {+}f_{uW,ii} \cwsq)$\\
& $--+0$ & $-\frac{ g\, \nc}{\sqrt{2}} f_{uW,ii}$\\
& $++-0$ & $-\frac{ g\,  \nc}{\sqrt{2}} f_{dW,ii}$\\
& $++0-$ & $-\frac{ g\,   \nc }{\sqrt{2}\cw}(f_{dB,ii} \swsq -f_{dW,ii} \cwsq)$\\
\hline
$ d_i  \bar{u_i}\to W^-A$
& $--0+$ & $-\frac{ g\, \sw  \,\nc}{\sqrt{2}}(f_{uB,ii} {-}f_{uW,ii})$\\
& $++0-$ & $+\frac{ g\, \sw  \,\nc}{\sqrt{2}}(f_{dB,ii}+f_{dW,ii})$\\
\hline 					
\end{tabular}
}
\caption{Unitarity violating   scattering  amplitudes 
  $\mathcal{M}({f_1}_{\sigma_1}{\bar{f_2} }_{\sigma_2} \to
  {V_3}_{\lambda_3}{V_4}_{\lambda_4})$ induced by the  operators
  in Eq.~(\ref{eq:dipole}).}
  \label{tab:DipoleAmp}
\end{table}

Assuming that the Wilson coefficients of the dipole operators are
generation independent, we can obtain, using
Table~\ref{tab:DipoleAmp}, the following unitarity bounds:
\begin{eqnarray}
\frac{1}{\sqrt{\Ng}}
 |{\displaystyle \sum_{i=1}^{\Ng}} e^-_{-,i}\,e^+_{-,i}\rangle
 \rightarrow W^+_0 W^-_+ 
 &\Rightarrow&
 \left|\frac{f_{eW}  }{\Lambda^2} s \right|\leq 49 \nonumber
 \\
\frac{1}{\sqrt{\Ng}}
 |{\displaystyle \sum_{i=1}^{\Ng}} e^-_{+,i}\,\bar\nu_{+,i}\rangle
\rightarrow W^+_{{0}} A_- 
&\Rightarrow& \left|\frac{(f_{eW}+f_{eB}) }{\Lambda^2} s \right|
\leq 144 \Rightarrow
\left|\frac{f_{eB} }{\Lambda^2} s \right|\leq 193\nonumber
\\
\frac{1}{\sqrt{\Ng\nc}}
|{\displaystyle \sum_{i=1}^{\Ng}}{\displaystyle \sum_{a=1}^{\nc}}
u^a_{-,i}\,\bar u^a_{-,i}\rangle \rightarrow W^+_+ W^-_0  
&\Rightarrow&
 \left|\frac{f_{uW}  }{\Lambda^2} s \right|\leq 28 \label{eq:magamp}
 \\
\frac{1}{\sqrt{\Ng\nc}}
|{\displaystyle \sum_{i=1}^{\Ng}} {\displaystyle \sum_{a=1}^{\nc}}
d^a_{+,i}\,\bar u^a_{+,i}\rangle \rightarrow W^-_0 A_+ 
&\Rightarrow& \left|\frac{(f_{uW}-f_{uB}) }{\Lambda^2} s \right| \leq 83 
\Rightarrow \left|\frac{f_{{uB}} }{\Lambda^2} s \right|\leq 111 \nonumber
\\
\frac{1}{\sqrt{\Ng\nc}}
|{\displaystyle \sum_{i=1}^{\Ng}} {\displaystyle \sum_{a=1}^{\nc}}
{d}^a_{-,i}\,\bar{{d}}^a_{-,i}\rangle \rightarrow W^+_0 W^-_+ 
&\Rightarrow&
 \left|\frac{f_{{dW}} }{\Lambda^2} s \right|\leq 28 \nonumber
 \\
 \frac{1}{\sqrt{\Ng\nc}}
 |{\displaystyle \sum_{a=1}^{\nc}}
 {\displaystyle \sum_{i=1}^{\Ng}} {d}^a_{+,i}\,\bar{{u}}^a_{+,i}\rangle
\rightarrow W^{{-}}_{{0}} A_- 
&\Rightarrow& \left|\frac{(f_{{dW}}+f_{{dB}}) }{\Lambda^2} s \right|
\leq 83 \Rightarrow
\left|\frac{f_{{dB}} }{\Lambda^2} s \right| \leq 111 \nonumber
\end{eqnarray}

Dropping the generation independence hypothesis for the dipole operators, 
we can use the same set of amplitudes as in
Eq.~(\ref{eq:magamp}) but now without summing over generations. In this case
the partial--wave unitarity constraints read:
\begin{eqnarray}
  \left|\frac{f_{eW,ii}  }{\Lambda^2} s \right|\leq 85\,,
  & & \left|\frac{f_{eB,ii}  }{\Lambda^2} s \right|\leq 334\, \\
  \left|\frac{f_{qW,ii} }{\Lambda^2} s \right|\leq 49\,,
  & & \left|\frac{f_{qB,ii} }{\Lambda^2} s \right|\leq 193\, 
\end{eqnarray}
where the last line applies for $q=u,d$. \smallskip

Because they flip the fermion chirality these operators do not
interfere at tree--level with the SM amplitudes and also generically
they are expected to be suppressed by the fermion Yukawa\footnote{ 
Bounds on the CP conserving coefficients of the dipole
operators for light fermions can be obtained, in principle, from their
tree-level  contribution to the corresponding anomalous magnetic moment,
but for the light quark dipole operators these are hard to extract in a model
independent way, therefore, being subject to large uncertainties.
For leptons, the current $g-2$ bounds~\cite{Giudice:2012ms} indicate that in
the absence of cancellations between the ${\cal O}_{eW}$
and ${\cal O}_{eB}$ contributions, for operators involving electrons or muons
unitarity is guaranteed  well beyond LHC energies.}. In this
case only the operators involving the top quark can be sizable.
There is an extensive study of the top quark
  properties at the LHC which includes the operators ${\cal O}_{uW}$
  and ${\cal O}_{uB}$; see, for instance, Ref.~\cite{Buckley:2015nca}
  and references therein.  In particular the measurement of W--boson
  helicity in top--quark decays~\cite{Aaltonen:2012rz} give us
  direct access to $f_{uW}/\Lambda^2$. Using the global fit to the top
  quarks properties in Ref.~\cite{Buckley:2015nca}
  ($|f_{uW}/\Lambda^2| < 3.8$ TeV$^{-2}$) we estimate that the
  operator ${\cal O}_{uW}$ does not lead to perturbative unitarity
  violation in top--quark processes at the LHC for maximum
  center--of--mass energies up to 2.7 TeV. \smallskip

\bibliography{references}
\end{document}